\documentclass[11pt, letterpaper]{article} 
\usepackage[margin=1in]{geometry}

\usepackage{natbib}
\usepackage{hyperref}
\hypersetup{colorlinks,citecolor=blue,linkcolor=cyan,urlcolor=blue}
\usepackage{color}

\usepackage{algorithmic}
\usepackage{algorithm}

\usepackage{amsmath}
\usepackage{amsfonts}
\usepackage{multirow}
\usepackage{multicol}   

\usepackage{footnote}
\makesavenoteenv{tabular}
\usepackage{caption}
\usepackage{subcaption}
\usepackage{graphicx}

\usepackage{epstopdf}

\usepackage{amsthm}
\newcommand\norm[1]{\left\lVert#1\right\rVert}

\title{\textbf{Reduced-Dimensional Reinforcement Learning Control using Singular Perturbation Approximations}}

\author{Sayak Mukherjee, He Bai, and Aranya Chakrabortty\thanks{Sayak Mukherjee and Aranya Chakrabortty are with Electrical and Computer Engineering Department, North Carolina State University, USA, and He Bai is with Mechanical and Aerospace Engineering Department, Oklahoma State University, USA.   Email: {\tt smukher8@ncsu.edu.}}}   
\date{\vspace{-5ex}}
\begin{document}
\maketitle

\begin{abstract}

We present a set of model-free, reduced-dimensional reinforcement learning (RL) based optimal control designs for  linear time-invariant singularly perturbed (SP) systems. We first present a state-feedback and output-feedback based RL control design for a generic SP system with unknown state and input matrices. We take advantage of the underlying time-scale separation property of the plant to learn a linear quadratic regulator (LQR) for only its slow dynamics, thereby saving significant amount of learning time compared to the conventional full-dimensional RL controller. We analyze the sub-optimality of the design using SP approximation theorems, and provide sufficient conditions for closed-loop stability. Thereafter, we extend both designs to clustered multi-agent consensus networks, where the SP property reflects through clustering. We develop both centralized and cluster-wise block-decentralized RL controllers for such networks, in reduced dimensions. We demonstrate the details of the implementation of these controllers using simulations of relevant numerical examples, and compare them with conventional RL designs to show the computational benefits of our approach.  
\vspace{-.5 cm}
\end{abstract}  

\section{Introduction}

Reinforcement Learning (RL), originally introduced in the artificial intelligence community \citep{barto}, has recently seen a resurgence in optimal control of dynamical systems through a variety of papers such as \citet{vrabie1,lewis_mag,jiang1,nonlin2,nonlin1,V17} using solution techniques such as adaptive dynamic programming (ADP),  actor-critic methods, Q-learning, etc. Curse of dimensionality, however, continues to be an ongoing debate for all of these RL-based control designs. Depending on the size and complexity of the plant, it may take an unacceptably long amount of time to even start the initialization step of RL, let alone control. {Our goal in this paper is to counteract this problem by exploiting certain physical characteristics of the plant dynamics that allow for model reduction so that learning only a reduced-dimensional controller is sufficient for stabilizing the full-dimensional plant. The specific property that we study is singular perturbation (SP). We consider plants whose dynamics are separated into two time-scales. Traditionally, SP theory has been used for model reduction (\citet{SPreduction, chow1985}), and control (\citet{chowslowfast}) of large-scale systems, but only by using full knowledge of the original plant model. Its extension to model-free control using RL has not been addressed. To bridge this gap, we present several sets of RL-based control designs where we exploit the underlying SP property of the plant to learn a controller for only its dominant slow time-scale dynamics, thereby saving significant amount of learning time. We provide  sub-optimality and stability results for the resulting closed-loop system.}

The main contributions are as follows. Three distinct RL control designs for singularly perturbed systems are presented. The first design assumes that the slow state variable is either directly measurable, or can be constructed from the measurements of the full state vector. Using this assumption, we develop a modified ADP algorithm which learns a reduced-dimensional RL controller using only feedback from the slow state variables. The controller is shown to guarantee closed-loop stability of the full-dimensional system if the fast dynamics are stable. The second design extends this algorithm to output feedback control using a neuro-adaptive state estimator \citep{obs}. The estimation of full-dimensional states is essential for our design to extract the slow states, in contrast to the time-shifted discrete-time output-feedback designs like \cite{obs_discrete} that uses a combination of inputs and outputs in the control law. The third design shows the relevance of these two designs to SP models of multi-agent consensus networks where time-scale separation arises due to clustering of the network nodes. Along with a centralized design, a variant is proposed that imposes a block-diagonal structure on the RL controller to facilitate its implementation. Numerical results show that our approach saves significant amount of learning time than the conventional RL while still maintaining a modest closed-loop performance. All the designs are described by implementable algorithms together with theoretical guarantees.


The first design has been presented as a preliminary result in our recent conference paper \citet{sayak_cdc}. The second design, however, is completely new. The multi-agent RL controllers, which were presented only for scalar dynamics in \citet{sayak_cdc, sayak_acc}, are now extended to  vector-dimensional states. Moreover, unlike prior results, the consensus model here is more generic as we allow each node to have self dynamics. The simulation examples presented in Section $7$ are much larger-dimensional than in \citet{sayak_cdc} to demonstrate the numerical benefits of the designs.

The rest of the paper is organized as follows. The state-feedback and output-feedback RL design problems are formulated in Section $2$, followed by their respective solutions and stability analyses in Sections 3 and 4. Section $5$ and $6$ interprets these designs to multi-agent consensus networks with node clustering, presenting both centralized and block-decentralized RL. Numerical simulations are shown in Section $7$. Concluding remarks are provided in Section $8$. Proofs of theorems and lemmas are presented in the Appendix.
\par
\textbf{Notations:}  $\mathbb{RH}_{\infty}$ is the set of all proper, real and rational stable transfer matrices; $\otimes$ denotes Kronecker product; $diag(m)$ is a diagonal matrix with vector $m$ on its principal diagonal; $\bf{1_n}$ denotes a column vector of size $n$ with all ones; $\cup$ denotes union operation of sets; $blkdiag(m_1,\dots,m_n)$ denotes a block-diagonal matrix with $m_1,\dots,m_n$ as its block diagonal elements; $|M|$ denotes the cardinality of set $M$; $\norm{.}$ denotes Euclidean norm of a vector and Frobenius norm of a matrix unless mentioned otherwise. 
\vspace{-.4 cm}
\section{Problem Formulation}
Consider a linear time-invariant (LTI) system 
\begin{align}
\label{eq:statecompact1}
& \dot{x} = Ax + Bu, \;\; x(0)=x_0,\;
 q=\mathcal C x,
\end{align}
 where, $x \in \mathbb{R}^n$ is the state, $u \in \mathbb{R}^m$ is the control input, and $q \in \mathbb{R}^p $ is the output. We assume that the matrices $A$ and $B$ are unknown, although the values of $n$, $m$ and $p$ are known. The following assumption is made. 
 
 \par
\noindent \textit{Assumption 1:} The system \eqref{eq:statecompact1} exhibits a singular perturbation property, i.e., there exist a small parameter $1 \gg \epsilon > 0$ and a similarity transform $\mathcal{T} = [T^T \; G^T]^T$ such that by defining $y \in \mathbb{R}^r$ and  $z \in \mathbb{R}^{n-r}$ as 
\begin{align}\label{similarity}
&\begin{bmatrix}
y \\ z
\end{bmatrix} = \mathcal{T}x = \begin{bmatrix}
T \\ G
\end{bmatrix}x,
\end{align} 
the state-variable model \eqref{eq:statecompact1} can be rewritten as
\begin{subequations}
\label{eq:SP}
\begin{align}
&\dot{y} = A_{11}y + A_{12}z + B_1u, \; \;y(0)=Tx_0=y_0, \\
 & \epsilon\dot{z} = A_{21}y + A_{22}z + B_2u,\; z(0)=Gx_0=z_0,\\
  & q=  \mathcal C \mathcal T^{-1}\begin{bmatrix}
y \\ z
\end{bmatrix} = C\begin{bmatrix}
y \\ z
\end{bmatrix}.
\end{align}
\end{subequations} \par
In the transformed model \eqref{eq:SP}, $y(t)$ represents the {\it slow} states and $z(t)$ represents the {\it fast} states. Since $A$ and $B$ are unknown, the matrices $A_{11}, A_{12}, A_{21}, A_{22}$, $B_1$ and $B_2$ are unknown as well. 

\vspace{-.2 cm}

\subsection{Problem Statement for State-Feedback RL}
\textbf{P1.} \textit{Learn} a control gain $K \in \mathbb{R}^{m \times r}$ for the singularly perturbed system \eqref{eq:SP} without knowing the model using online measurements of $u(t)$ and $y(t)$ such that 
\begin{equation}
    u(t) = -Ky(t) = -KTx(t) \label{spu}
\end{equation} minimizes 
\begin{align}\label{state_J} &J(y(0);u)= \int_0^{\infty} (y^T Q y + u^T R u )dt, \\
&\mbox{s.t.} \;\; A- BKT \in \mathbb{RH}_{\infty}.
\end{align}
We assume $(A,B)$ to be stabilizable. We consider $y(t)$ to be directly measurable, or $x(t)$ to be measurable (i.e. $\mathcal C=I$) and $T$ to be known so that $y(t)$ can be computed at all time $t$. This is not a restrictive assumption as in many SP systems the identity of the slow and fast states are often known a priori \citep{Khalilcontrol} even if the model is unknown. If the system is explicitly represented in form (2), then $ \mathcal{T} = I$, and we assume that the slow variable $y(t)$ is available. 
The benefit of using $y(t)$ as the feedback variable is that  one has to learn only a $(m\times r)$ matrix instead of a $(m \times n)$ matrix if full state feedback $x(t)$ was used. This will improve the learning time, especially if $r \ll n$. Before proceeding with the control design, we make the following assumption.
\par
\noindent \textit{Assumption 2:} $A_{22}$ in (3b) is Hurwitz. \par
This assumption means that the fast dynamics of \eqref{eq:SP} are stable, which allows us to skip feeding back $z(t)$ in \eqref{spu}. 
\vspace{-.2 cm}
\subsection{Problem Statement for Output Feedback RL}
\textbf{P2.} Considering that $q(t)$ is measured and $\mathcal C$ is known, but $A$ and $B$ are both unknown in \eqref{eq:statecompact1}, \textit{estimate} the states $\hat{y}(t),\hat{z}(t)$ (or, equivalently estimate $\hat{x}(t)$ and compute $\hat{y}(t)=T\hat{x}(t)$ assuming that $T$ is known), \textit{learn} a controller $K \in \mathbb{R}^{m\times r}$ using online measurements of $q(t)$ and $u(t)$ such that  
\begin{equation}
    u = -K\hat{y} = -KT\hat{x}
\end{equation} 
minimizes 
\color{black}
\begin{align}\label{output_J}
J(y(0);u)= \int_0^{\infty} (y^T Q y + u^T R u )dt.
\end{align}
\color{black}
We assume $(A,B)$ to be stabilizable, and $(A,C)$ to be detectable. Our approach would be to estimate the slow states $\hat{y}(t)$ without knowing $(A,\,B)$ using an observer employing a neural structure that does not require exact information of the state dynamics, and then using $u(t)$ and $\hat{y}(t)$ to learn the controller $K$ using adaptive dynamic programming.

\par
We present the solutions for $\bf{P1}$ and $\bf{P2}$ with associated stability proofs in the following two respective sections. 
\vspace{-.25 cm}
\section{Reduced-dimensional State Feedback RL}
Following \citet{khalil}, the \textit{reduced} slow subsystem of \eqref{eq:SP} can be defined by substituting $\epsilon=0$, resulting in
\begin{align}
\label{eq:slowsubsystem}
&\dot{y}_s = A_s y_s + B_s u_s, \;\; y_s(0)= y(0), \;\;
u=u_s +u_f,
\end{align}
where $A_s = A_{11} - A_{12}A_{22}^{-1}A_{21}$ and $B_s = B_1 - A_{12}A_{22}^{-1}B_{2}$.
Since our intent is to only use the slow variable for  feedback, we substitute the fast control input $u_f=0$, and the slow control input $u_s=u$. If the controller were to use $y_s(t)$ for feedback then it would find $u = -\bar{K}y_s(t)$ to solve: 
\begin{align}\label{J_red}
\text{minimize} \;\; &\bar{J}(y_s(0);u)= \int_0^{\infty} (y_s^T Q y_s + u^T R u )dt, \\
&\mbox{s.t.} \;\; A_s - B_s\bar{K} \in \mathbb{RH}_{\infty}.
\end{align}
The optimal solution for the above problem is given by the following algebraic Riccati equation (ARE):
\begin{align}
&A_s^T\bar{P} + \bar{P}A_s + Q -\bar{P}B_sR^{-1} B_s^T\bar{P} = 0,
\bar{K} = R^{-1}B_s^T\bar{P},\nonumber
\end{align}
\par 
where $\bar{P} = \bar{P}^T \succ 0$. If $A_s$ and $B_s$ are unknown, then the RL controller $\bar{K}$ can be learned using measurements of $y_s(t)$ and of an exploration input $u(t)=u_0(t)$ by the ADP algorithm presented in \cite{jiang_book}, which is an iterative version of Kleinman's algorithm \citet{kleinman}. The control policy $u_0(t)$ must be persistently exciting, and can be chosen arbitrarily as long as the system states remain bounded. For example, one choice of $u_0$ is a sum of sinusoidal signals.

In reality, however, $y_s$ is not accessible as $\epsilon \neq 0$. We, therefore, recall the following theorem from \citet{chowslowfast}, which will allow us to replace $y_s(t)$ with $y(t)$ in the learning algorithm. 
\par
\noindent \textbf{Theorem 1 \citep{chowslowfast,khalil}:}
\textit{Consider the two systems \eqref{eq:SP} and \eqref{eq:slowsubsystem}. There exists $0<\epsilon^* \ll 1$ such that for all $0< \epsilon \leq \epsilon^*$, the trajectories $y(t)$ and $y_s(t)$ satisfy uniformly for $t \in [0,t_1]$}
\begin{align}
y(t) = y_s (t) + O(\epsilon).
\end{align}
Algorithm 1 shows how the controller $K$ is learned using $y$ and $u_0$, based on \citet{jiang1}.

\begin{algorithm}[t]
\footnotesize
\caption{SP-RL using slow dynamics}
\textbf{Input:} Measurements of $y(t)$ and $u_0(t)$\\
\textbf{Step 1 - Data storage:}
\textit{Store} data (i.e., $y(t)$ and $u_0(t)$) for sufficiently large uniformly sampled time instants $(t_1,t_2,\cdots,t_l)$, and  \textit{construct} the following matrices:
\vspace{-.47 cm}
\begin{align} 
& \hspace{-.3 cm} \delta_{yy} = \begin{bmatrix}
y \otimes y |_{t_1}^{t_1+T} ,& \cdots &, y \otimes y |_{t_l}^{t_l+T}
\end{bmatrix}^T,\\
& \hspace{-.3 cm} I_{yy} = \begin{bmatrix}
\int_{t_1}^{t_1+T}(y \otimes y) d\tau ,& \cdots &, \int_{t_l}^{t_l+T} (y \otimes y) d\tau \\
\end{bmatrix} ^T,\\
& \hspace{-.3 cm} I_{yu_0} = \begin{bmatrix}
\int_{t_1}^{t_1+T}(y \otimes u_0) d\tau ,& \cdots & ,\int_{t_l}^{t_l+T} (y \otimes u_0) d\tau \\
\end{bmatrix} ^T,
\end{align} 
such that rank($I_{yy} \;\; I_{yu_0}) = r(r+1)/2 + rm$ satisfies. 

\textbf{Step 2 - Controller update:}
Starting with a stabilizing $K_0$, \textit{solve} for $K$ iteratively ($k=0,1,\cdots$) following the update equation:
\vspace{-.6 cm}
\begin{align}\label{eq:updateA1}
\underbrace{\begin{bmatrix}
\delta_{yy}  -2I_{yy}(I_r \otimes K_k^TR)  -2I_{yu_0}(I_r \otimes R)
\end{bmatrix}}_{\Theta_k}\begin{bmatrix}
vec(P_k) \\ vec(K_{k+1})
\end{bmatrix}  =\underbrace{-I_{yy}vec(Q_k)}_{\Phi_k}.
\end{align}
The stopping criterion for this update is $\norm{P_k - P_{k-1}} < \gamma$, where $\gamma$ is a chosen small positive threshold. \\
\textbf{Step 3 - Applying control:} After $P$ and $K$ converge, remove $u_0$ and apply $u=-Ky$.
\end{algorithm} 
\normalsize
The condition rank($\Theta_k$) = $r(r+1)/2 + rm$ can be satisfied, for example, by utilizing data from at least twice as many sampling intervals as the number of unknowns. We next provide the analytical guarantees of Algorithm $1$ related to the SP-based approximations.
\vspace{-.3 cm}
\subsection{Sub-optimality and Stability Analysis}
The optimal controller parameters $P,K$ can be written as
$
P=\bar{P} + \Delta P ,
K=\bar{K} + \Delta K, 
$
where $\bar{P},\bar{K}$ are the optimal solutions if $y_s(t)$ were available for design, and $\Delta P, \Delta K$ are matrix perturbations resulting from the fact that $\epsilon \neq 0$. The following theorem establishes the  sub-optimality of the learned controller using $y(t)$. 

\noindent \textbf{Theorem 2:}  \textit{Assuming $||y_s(t)||$ and $||u_0(t)||$ are bounded for a finite time $t \in [0,t_1]$, the solutions of Algorithm 1 are given by $P=\bar{P} + O(\epsilon)$, $K=\bar{K} + O(\epsilon)$, and $ J =\bar{J} + O(\epsilon)$.}

\noindent \textit{Proof:} See theorems $2$ and $3$ in \citet{sayak_cdc}.\par
Theorem $2$ shows that the controller obtained from Algorithm 1 is $O(\epsilon)$ close to that obtained from the ideal design using the actual slow variables. Next, we analyze how this perturbation affects the optimal objective. The next theorem provides a sufficient condition that is required to achieve asymptotic stability for the $(k+1)^{th}$ iteration of Algorithm 1 assuming that the control policy at the $k^{th}$ iteration stabilizes \eqref{eq:SP}.

\noindent \textbf{Theorem 3:} 
\textit{Assume that the control policy $u = -K_ky$ at the $k^{th}$ iteration asymptotically stabilizes \eqref{eq:SP}. Consider $R \succ 0$ and $Q \succ 0$ with $ \lambda_{min}(Q)$ sufficiently large. Then the control policy at the $(k+1)^{th}$ iteration given by $u =- K_{k+1}y$ is asymptotically stabilizing for \eqref{eq:SP}.} \qed

\noindent \textit{Proof:} Please see Theorem $4$ in \citet{sayak_cdc}.

\noindent \textbf{Remark 1: (Design trade-off)} The proof of Theorem 3 is based on Lyapunov function based stability analysis, where $Q$ compensates for the error due to $O(\epsilon)$ approximation of the fast dynamics such that $Q - O(\epsilon) \succ 0$. This translates to the requirement of a sufficiently large $\lambda_{min}(Q)$. In practice, one can start the off-policy RL iteration in a computing platform after gathering sufficient data with a considerable $Q \succ 0$, and if that is found to be not stabilizing then tune $Q$ until the states are bounded.
\vspace{-.4 cm}
\section{Reduced-Dimensional Output Feedback RL}
We next address the RL design when the full state information is not available. We start by considering the generic system \eqref{eq:SP}, and then design an observer to estimate the state $x$ as $\hat{x}(t) = [\hat{y}(t);\hat{z}(t)]$. As $T$ is known, the slow state can be estimated as $\hat{y}(t) = T\hat{x}(t)$. The idea then is to simply replace $y(t)$ by $\hat{y}(t)$ in Algorithm 1.   
Algorithm $2$ shows the steps for this output feedback RL-based design. 
In Section $4.2$ we will present one such observer which can estimate $x$  without having a proper knowledge about the model \eqref{eq:SP}. Before that, we first analyze the stability properties of the output feedback design. 
\begin{algorithm}
\footnotesize
\caption{Output feedback ADP/RL}
\textbf{Input:} Measurements of $\hat{y}(t)$ and $u_0(t)$\\
\textbf{Step 1 - Data storage:} Construct the matrices $\delta_{\hat{y}\hat{y}},I_{\hat{y}\hat{y}},I_{\hat{y}u_0}$ with similar structures as $\delta_{yy},I_{yy},I_{yu_0}$ respectively but with $y(t)$ replaced by $\hat{y}(t)$.\\
\textbf{Step 2 - Controller update:}
Following Step $2$ of Algorithm $1$, \textit{update} the control gains as: 
\begin{align}
&\underbrace{\begin{bmatrix}
 \delta_{\hat{y}\hat{y}}  -2I_{\hat{y}\hat{y}}(I_r \otimes K_k^TR) -2I_{\hat{y}u_0}(I_r \otimes R)
\end{bmatrix}}_{\hat{\Theta}_k}\begin{bmatrix}
vec(P_k) \\ vec(K_{k+1})
\end{bmatrix} = \underbrace{-I_{\hat{y}\hat{y}}vec(Q_k)}_{\hat{\Phi}_k}.
\end{align}
The stopping criterion for this update is $\norm{P_k - P_{k-1}} < \gamma_1$, where $\gamma_1$ is a chosen small positive threshold.\\
\textbf{Step 3 - Applying control:} Remove $u_0$ and apply  $\tilde{u} = -K\hat{y}$. 
\end{algorithm}
\vspace{-.35 cm}
\subsection{Sub-optimality and Stability Analysis}
\textbf{Lemma 1:} Define $e(t)=x(t)-\hat{x}(t)$. If $e$ is uniformly ultimately bounded (UUB) with a bound $b$ for all $t \geq t_0 +T_1$ for some initial time $t_0$, then there exists positive constants $\epsilon^{*}$ and $k$ such that for all $0 < \epsilon \leq  \epsilon^{*}$
\vspace{-.3 cm}
\begin{align}
||\hat{y}(t) - y_s(t) || \leq \bar{k}|\epsilon| + b := c(\epsilon,b)
\end{align}
holds uniformly for $t \in [t_2,t_1]$.\\
\textit{Proof:}
Since $e(t)$ is UUB, there exists positive constants $b$ and $\hat{b}$, independent of $t_0 \geq 0,$ and for every $a \in (0,\hat{b})$, there exists $T_1=T_1(a,b)$, independent of $t_0$, such that $||\hat{y}(t_0) - y(t_0)|| \leq a$, which implies that
\begin{equation} ||\hat{y}(t) - y(t)|| \leq b, \; \forall t \geq t_0 + T_1 :=t_2. \label{int1}
\end{equation}
From Theorem $1$, it follows that there exist positive constants $k$ and $p$ such that,
\begin{align}
\hspace{-.4 cm}||y(t) - y_s(t)|| \leq \bar{k}|\epsilon | \;\;\;\;  \forall t \in [t_0,t_1], t_1>t_2, \forall |\epsilon| < p. \label{int2}
\end{align}
Combining \eqref{int1} and \eqref{int2}, for $ t \in [t_2, t_1]$ we have
\begin{align}
& ||\hat{y}(t) - y_s(t) || \leq \bar{k}|\epsilon| + b := c(\epsilon,b).
\end{align}
This completes the proof.\qed \\
\textbf{Corollary 2:} If $e(t) = O(\epsilon)$ for $t \in [t_2,t_1]$, then $\hat{y}(t) = y_s(t) + O(\epsilon).$

\noindent \textit{Proof:} The proof directly follows from Lemma $1$. \qed \\

We know that if $y_s(t)$ were available for feedback then $\bar{P},\bar{K}$ would be the optimal solutions. However, due to the state estimation error bound $b$ and the singular perturbation error $O(\epsilon)$, the actual solutions are given as $P=\bar{P} +\Delta P$, $ K=\bar{K} + \Delta K$,
where $\Delta P$ and $\Delta K$ are matrix perturbations resulting from non-ideal feedback.

\noindent \textbf{Proposition 1: } Perturbations $\Delta P, \Delta K$ are bounded, i.e., there exist two positive constants $\rho,\, \rho_1$, dependent on $b$ and $\epsilon$, such that $\norm{\Delta P}\leq \rho, \norm{\Delta K} \leq \rho_1$. Moreover, if $e(t) = O(\epsilon)$ for $t \in [t_2,t_1]$, then we will recover $P= \bar{P} + O(\epsilon), K=\bar{K} + O(\epsilon)$.

\noindent\textit{Proof:} 
Please see Appendix A. 
\color{black}

If $e(t) $ can be made sufficiently small by proper tuning of the observer gain then we would recover the design characteristics of Algorithm 1. To this end, we present the following stability result.

\noindent \textbf{Theorem 4:} Assume that the control policy $u = -K_k \hat{y}$ is asymptotically stabilizing for the $k^{th}$ iteration in Step 2 of Algorithm 2. Then, there exist sufficiently small $b^{*},$ and $0<\epsilon^* \ll 1$ such that for $b \leq b^*, 0 < \epsilon \leq \epsilon^*$, with $Q \succ 0, R \succ 0$, $u=-K_{k+1}\hat{y}$ will asymptotically stabilize \eqref{eq:SP} at the $(k+1)^{th}$ iteration.\par
\noindent \textit{Proof:} Please see Appendix B.\\
As shown in Appendix B, the estimation error enters the closed-loop system as an exogenous disturbance. Since $K_{k+1}$ is stabilizing, the states  converge to a neighborhood of the origin for sufficiently small $b^*$ and $\epsilon^*$. Note that the designer does not need the explicit knowledge of $\epsilon^*$, and can simply assume a strong time-scale separation in the plant dynamics resulting in a small enough $\epsilon$.

\noindent \textbf{Remark 2:} The convergence of the observer dynamics and that of the RL iterations are handled sequentially. The observer is used to gather sufficient amount of data samples to meet the rank condition $\mbox{rank}(\hat{\Theta}_k) =r(r+1)/2 + rm$, after which the  control gain is computed iteratively. $\hat{\Theta}_k$ has same structure as $\Theta_k$ but with $y(t)$ replaced by $\hat{y}(t)$. The designer may start gathering data samples after a few initial time-steps over which the observer may have converged close to its steady-state. The observer is  designed  to  achieve  fast convergence, as discussed next. The state  estimation  error  that may be present  in the observer output has been taken into consideration in the sub-optimality and the stability analysis, as discussed in Proposition 1 and Theorem 4.
\vspace{-.3 cm}
\subsection{Neuro-adaptive Observer}
A candidate observer to estimate $\hat{y}(t)$ without knowing $(A,\,B)$ is the neuro-adaptive observer proposed in \citet{obs}. The observer employs a neural network structure to account for the lack of dynamic model information. This observer guarantees  boundedness of $e(t)$, which, with proper tuning, can also be made arbitrarily small. We next recall the mechanism of this observer. We rewrite \eqref{eq:statecompact1} as
\begin{align}\label{obs1}
&\dot{x} = \hat{A}x + \underbrace{(Ax - \hat{A}x) + Bu}_{g(x,u) },
\; q=\mathcal{C}x,
\end{align}
where $\hat{A}$ is a Hurwitz matrix, and $(\mathcal{C},\hat{A})$ is observable. We do not have proper knowledge about $g(x,u)$, and a neural network (NN) with sufficiently large number of neurons can approximate $g(x,u)$, as $ g(x,u) = W\sigma (V\bar{x}) + \eta(x)$. Here, $\bar{x} = [x,u]$, while $\sigma(.)$ and $\eta(x)$ are the activation function and the bounded NN approximation error, respectively. $W$ and $V$ are the ideal fixed NN weights. We choose $G$ such that $ \hat{A}-G\mathcal{C}$ is Hurwitz. The observer dynamics follow as
\begin{align}\label{observer}
& \dot{\hat{x}} = \hat{A} \hat{x} + \underbrace{g(\hat{x},u)}_{ = \hat{W}\sigma (\hat{V} \hat{\bar{x}})} + G(q - \mathcal{C}\hat{x}),\;
\hat{q} = \mathcal{C}\hat{x},
\vspace{-.4 cm}
\end{align}
 where $\hat{W},\,\hat{V}$ are neural network weights when driven by $\hat{x}$, and are updated 
based on the modified Back Propagation (BP) algorithm. The observer \eqref{observer} requires the knowledge of $\mathcal{C}$. Accordingly, we define the output error as $\tilde{q} =  q - \mathcal{C}\hat{x}$. The objective function is to minimize
$
 J = \frac{1}{2}(\tilde{q}^T\tilde{q}).
$
Following \citet{obs}, the update law follows from gradient descent as:
\begin{align}\label{updateW}
&\dot{\hat{W}} = - \eta_1 (\tilde{q}^T \mathcal{C} A_c^{-1})^T(\sigma(\hat{V}\hat{\bar{x}}))^T - \rho_1 ||\tilde{q}||\hat{W},\\
&\dot{\hat{V}} = - \eta_2 (\tilde{q}^T \mathcal{C} A_c^{-1} \hat{W}(I - \Lambda(\hat{V}\hat{\bar{x}})))^T \mbox{sgn}(\hat{\bar{x}})^T - \rho_2 ||\tilde{q}||\hat{V},\nonumber 
\end{align}
where, $\eta_1,\, \eta_2 > 0$ are learning rates and $\rho_1, \, \rho_2$ are small positive numbers. Considering $k$ neurons we have $ \Lambda(\hat{V}\hat{\bar{x}})) = diag(\sigma_i^2(\hat{V}_i \hat{\bar{x}})), i=1,2,..,k$, where sgn(.) is the sign function. The update law \eqref{updateW} depends on the knowledge of $\mathcal{C}$. This observer guarantees the following boundedness property.\\ 
\textbf{Theorem 5 \citep[Theorem 1]{obs}:} With the update law described as \eqref{updateW}, the state estimation error $\tilde{x} = x-\hat{x}$ and weight estimation errors $\tilde{W} = W-\hat{W}, \tilde{V}= V-\hat{V}$ are uniformly ultimately bounded (UUB). 

The size of the estimation error bound can be made  arbitrarily small by properly selecting the parameters and learning rates as shown in \citet{obs}. Selecting $\hat{A}$ to have fast eigenvalues will also keep the state estimation error small. 

\section{Applying to Clustered Multi-Agent Networks}

We next describe how SP-based RL designs can be applied for the control of clustered multi-agent consensus networks. Example of such networks abound in practice including power systems, robotic swarms, and biological networks. The LTI model of these networks can be brought into the standard SP form \eqref{eq:statecompact1} by exploiting the time-scale separation in its dynamics arising from the clustering of nodes. 
\vspace{-.35 cm}
\subsection{SP representation of clustered networks}
Consider a network of $n$ agents, where the dynamics of the $i^{th}$ agent is given by
\begin{align}
\label{eq:network}
\dot{x}_i = Fx_i+ \sum_{j \in \mathcal{N}_i} a_{ij} (x_{j} - x_{i}) + b_iu_i, 
\end{align} 
where $x_i \in \mathbb{R}^{s}$ is the state, $u_i \in \mathbb{R}^{p}$ is the input, and $\mathcal{N}_i$ denotes the set of agents that are connected to agent $i$, for $i =1,\dots n$. The connection graph between agents is assumed to be connected and time-invariant. The constants $a_{ij} = a_{ji} > 0$ denote the coupling strengths of the interaction between agents $i$ and $j$, and vice versa. The matrix $F \in \mathbb{R}^{s \times s}$ models the self-feedback of each node. The overall network model is written as 
\begin{align}
\label{eq:statecompact}
\dot{x} = Ax + Bu, \;\; x(0)=x_0,
\end{align} 
where, $x \in \mathbb{R}^{ns}$ is the vector of all agent states, $u \in \mathbb{R}^{ns}$ is the control input, $B= diag(b_1,\dots,b_n)$, $A = I_n\otimes F + L\otimes I_s$,  $L \in \mathbb{R}^{n \times n}$ being the weighted network Laplacian matrix satisfying $L\bf{1}_n = \bf{0}$. 

\noindent \textit{Assumption 3:} $F$ is marginally stable.   

\textcolor{black} {Let the agents be divided into $r$ non-empty, non-overlapping, distinct groups ${\mathcal{I}_1,\dots,\mathcal{I}_r}$ such that agents inside each group are {\it strongly} connected while the groups themselves are {\it weakly} connected. In other words, $a_{ij} \gg a_{pq}$ for any two agents $i$ and $j$ inside a group and any other two agents $p$ and $q$ in two different groups.  This type of clustering has been shown to induce a two-time scale behavior in the network dynamics of \eqref{eq:network}. Please see \cite{chow1985} for details. Fig. \ref{arch_central} shows an example of such a clustered dynamic network.} The clustered nature of the network helps decompose $L$ as $L = L^I + \epsilon L^E$, where $L^I$ is a block-diagonal matrix that represents the internal connections within each area, $L^E$ is a sparse matrix that represents the external connections, and $\epsilon$ is the singular perturbation parameter arising from the worst-case ratio of the coupling weights inside a cluster to that between the clusters. The slow and fast variables are defined as 
\begin{align}\label{similarity}
&\begin{bmatrix}
y \\ z
\end{bmatrix} = \begin{bmatrix}
T \\ G
\end{bmatrix}x,\;\;
 x = (U \;\; G^\dagger)\begin{bmatrix}
y \\ z
\end{bmatrix},
\end{align}
where, $T =T_1 \otimes I_s, G=G_1 \otimes I_s$. The definitions of $T_1 \in \mathbb{R}^{r\times n}$ and $G_1 \in \mathbb{R}^{(n-r)\times n}$ can be found in  \cite{chow1985}. Applying this transformation to \eqref{eq:statecompact}, and redefining the time-scale as $t_s = \epsilon t$, the following SP form is obtained:
\begin{subequations}\label{eq:sp_stan}
\label{eq:SP2}
\begin{align}
&\frac{dy}{dt_s} = A_{11}y + A_{12}z + B_1u, \\
 & \epsilon\frac{dz}{dt_s} = A_{21}y + A_{22}z + B_2u,
\end{align}
\end{subequations}
\vspace{-.4 cm}
\begin{align*}
   & A_{11} = T(L^E \otimes I_s)U + (I_r \otimes F)/\epsilon,
A_{12}= T(L^E \otimes I_s)G^{\dagger}, \\
&
A_{21} = G(L^E \otimes I_s)U, A_{22} = G(L^I \otimes I_s)G^\dagger + (I_{n-r} \otimes F) + \\ & \epsilon G(L^E \otimes I_s)G^\dagger,
B_1 = TB/\epsilon,B_2= GB.
\end{align*}
 The detailed derivation is shown Appendix C. 
\color{black}
All six matrices are assumed to be unknown.  Following Assumption 2, we assume that $A_{22}$ is Hurwitz. 
\vspace{-.35 cm}
\subsection{Projection of control to agents}

One important distinction between controlling the multi-agent system \eqref{eq:SP2} and a generic SP system \eqref{eq:SP} is that the control input $u$ for the former has a physical meaning in terms of each agent. Therefore, even if $u$ is designed using a reduced-dimensional controller, it must be actuated in its actual dimension. \textcolor{black}{One way to design $u(t)$ can be to use $u = M\tilde{u}$ where $\tilde{u} \in \mathbb{R}^{(rp)\times (rs)}$ is the actual control signal learned using ADP, and the matrix $M$ is a projection matrix of the form
$
M = blkdiag(M^{1},\dots,M^{r}),M^{i}=\bar{M}^i \otimes I_s, \bar{M}^i = \bf{1}_{|\mathcal{I}_i|}, 
$
 which projects the reduced-dimensional controller to the full-dimensional plant. The projection matrix $M$ is constructed by the designer with the assumption that the designer knows the cluster identity of each agent.} 
We assume $(A,BM)$ to be stabilizable. The same back-projection concept can be used for output feedback RL. 
\vspace{-.3 cm}
\section{Block-decentralized Multi-agent RL}
\begin{figure*}
        \begin{subfigure}[b]{1\textwidth}
                \centering
                
                \includegraphics[width=.8\linewidth, trim = 4 4 4 4,clip]{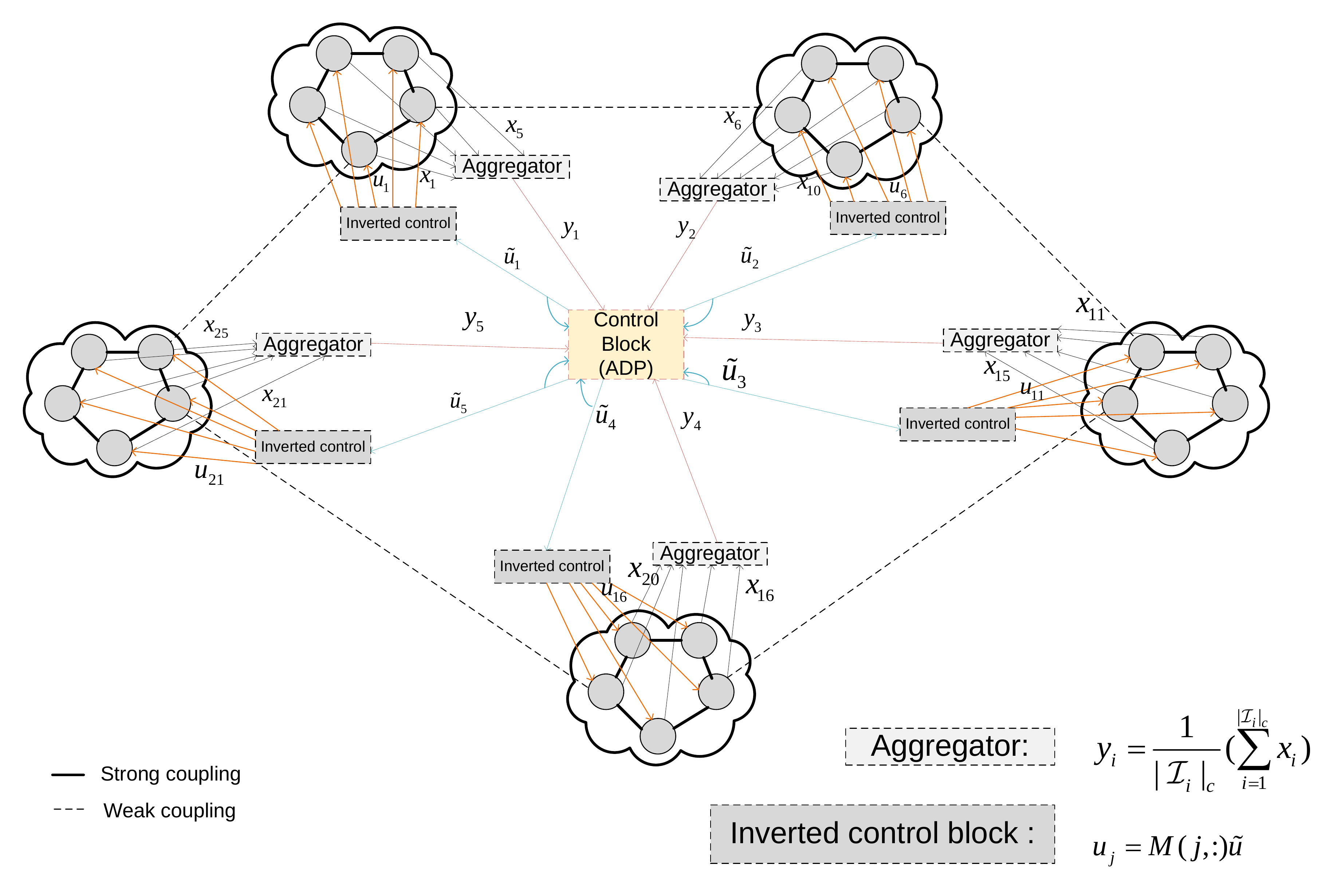}
                \caption{\small{Centralized control architecture for the clustered network}}
                \label{arch_central}
                
        \end{subfigure}
       \qquad
             \begin{subfigure}[b]{1\textwidth}
                \centering
                
                \includegraphics[width=.8\linewidth,  trim = 4 4 4 4,clip]{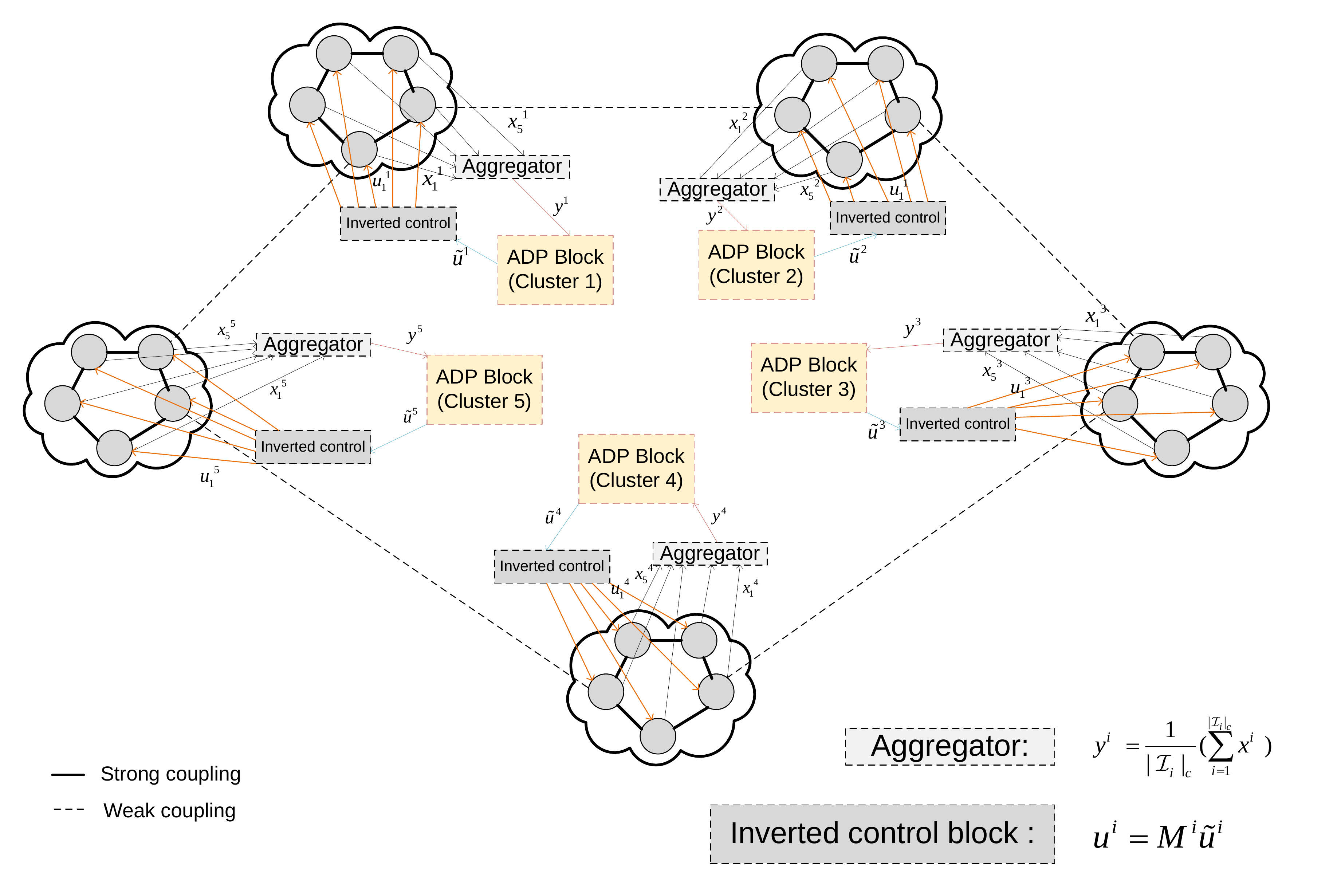}
                \caption{\small{Control architecture for the area-wise decentralized design}}
                \label{arch_decen}
                
        \end{subfigure}
        
\caption{\small{Centralized and block-decentralized control architectures }}
\label{fig:coherent}
\vspace{- .4 cm}
\end{figure*}

The controllers learned in Section $3$ and $4$ need to be computed in a centralized way. In this section we show that for the clustered consensus model \eqref{eq:SP2} the clustered nature of the system can also aid in learning a cluster-wise decentralized RL controller. Figs. \ref{arch_central},\ref{arch_decen} describe the centralized and block-decentralized architectures.
\vspace{-.35 cm}
\subsection{Cluster-wise representation}
Let the states of the agents in cluster $\alpha$ be denoted as $(x_{1}^\alpha,\,x_{2}^\alpha,\dots,x_{n_{\alpha}}^\alpha) \in \mathbb{R}^{n_\alpha s}.$ Following \cite{chow1985}, the transformation matrix $T$ in \eqref{similarity} is an averaging operation on the states of agents inside a cluster, 
which implies that the slow variable for the cluster $\alpha$ is
\vspace{-.45 cm}
\begin{align}\label{eq:y}
& y^{\alpha} = \frac{1}{n_{\alpha}}(x_{1}^{\alpha}+x_{2}^{\alpha}+\dots+x_{n_{\alpha}}^{\alpha}),\; \alpha=1,\dots,r,\\
& y=[y^1;y^2;\dots;y^r].\label{eq:y1}
\end{align}
For the cluster-wise decentralized design, the starting point is to consider the scenario if all clusters were decoupled from each other. We denote the states in cluster $\alpha$ in that scenario as $x_{d1}^\alpha,\,x_{d2}^\alpha,\dots,x_{dn_{\alpha}}^\alpha \in \mathbb{R}^{n_\alpha s} $, and the concatenated state vector considering all the clusters are denoted as $x_d$. For this decoupled scenario, $y_d^\alpha$ and $y_d$ are similarly defined following \eqref{eq:y} and \eqref{eq:y1}. Then we will have,
\vspace{-.35 cm}
\begin{align}\label{yd}
    &\dot{x}_d = (I_n \otimes F + L^I \otimes I_s)x_d + Bu,\\
    &\dot{y}_d = T\dot{x}_d= (T_1 \otimes I_s)(I_n \otimes F + L^I \otimes I_s)x_d + \tilde{B}_1u, \nonumber 
\end{align}
where $\tilde{B}_1 = TB$. As $x_d = U y_d + G^\dagger z_d$, \eqref{yd} is reduced to
\begin{align}
    \dot{y}_d = (I_r \otimes F)y_d + \tilde{B}_1 u. \label{dec}
\end{align}
The controller can be represented cluster-wise as
$
u =[u^1;\,u^2;\,\dots,\,u^r].
$ 
Using the projected controller discussed in Section 5.2,  we can design $u^\alpha(t)$ as
\begin{align}
& u^{\alpha} = M^{\alpha}\tilde{u}^{\alpha},\;
 M^{\alpha}=\bar{M}^\alpha \otimes I_s, \bar{M}^i = \bf{1}_{|\mathcal{I}_i|_c},
\end{align} 
where $\tilde{u}^{\alpha} $ is the controller learned in cluster $\alpha$, $\alpha = 1,\cdots,r$.
Taking a hint from the cluster-wise decentralized structure of $y_d$-dynamics in \eqref{dec}, we next state our design problem as follows.

\noindent \textbf{P3.} Consider the multi-agent consensus model \eqref{eq:statecompact} where $A$ and $B$ are unknown. \textit{Learn} a control gain $K^{\alpha}$ for every area $\alpha$, $\alpha = 1,\dots,r$, using $y^\alpha(t)$ and $\tilde{u}^\alpha (t)$ such that $u^{\alpha} = M^{\alpha}\tilde{u}^{\alpha} = -M^{\alpha} K^\alpha y^\alpha $ stabilizes the closed-loop system and minimizes the following individual cluster-wise objectives 
\vspace{-.35 cm}
\begin{align}
J^\alpha(y^\alpha (0);\tilde{u}^\alpha)= & \int_0^{\infty} (y^{\alpha T} Q^\alpha y^\alpha + \tilde{u}^{\alpha T} R^\alpha \tilde{u}^\alpha )dt,
\end{align}  
for $\alpha = 1,\dots,r$. We assume that $(A,BM)$ is stabilizable. 
\vspace{-.35 cm}
\subsection{RL Algorithm}

We exploit a different $O(\epsilon)$ separation existing between the trajectories of the actual average variable of an area and the same variable when the areas are decoupled. We start by providing a lemma proving how the actual average variable $y^\alpha$ is related to the decoupled average variable $y_d^\alpha$ for an area $\alpha$. 
\par
\noindent \textbf{Lemma 2:} \textit{The cluster-wise average variable $y^\alpha(t)$ and the decoupled average variable $y_d^\alpha(t)$ are related as,}
\vspace{-.35 cm}
\begin{align}
y^\alpha(t) = y_d^\alpha(t) + O(\epsilon), \forall t \in [0,t_1].
\end{align}
\noindent \textit{Proof:} 
The proof is shown in Appendix D. 
\color{black}
\par
We first consider the scenario when the clusters are decoupled. The average operation can be considered accordingly in $T$. The decoupled slow dynamics is given in \eqref{yd}.
The controller for area $\alpha$ uses the $y_{d}^{\alpha}(t)$ feedback and implements $\tilde{u}^\alpha = -\bar{K}^{\alpha}y_{d}^{\alpha}(t)$ so that the decoupled dynamics are stabilized and the following objective is minimized for area $\alpha$ with the ARE solution $\bar{P}^\alpha \succ 0$ and the optimal control gain $\bar{K}^\alpha$: 
\vspace{-.35 cm}
\begin{align}\label{Jbar}
& \hspace{-.3 cm} \bar{J}^{\alpha}(y_{d}^\alpha(0);\tilde{u}^\alpha(0))= \int_0^{\infty} (y_d^{\alpha T} Q^\alpha y_d^\alpha + \tilde{u}^{\alpha T} R^\alpha \tilde{u}^\alpha )dt.
\end{align}
\vspace{-.45 cm}
As the decoupled system is fictitious, based on Lemma 2, it is plausible to replace $y_d^\alpha(t)$ with $y^\alpha(t)$ in the learning algorithm and then follow the same procedure as the Kleinman's algorithm. The resulting algorithm is given in Algorithm $3$.
\footnotesize
\begin{algorithm}[] 
\footnotesize
\caption{  Cluster-wise Decentralized ADP} 
\label{alg1} 
\textbf{For} area $\alpha = 1,2,\dots,r$ \\
\textbf{Step 1:} Construct matrices $\delta_{y^\alpha y^\alpha },I_{y^\alpha y^\alpha },I_{y^\alpha u_0^\alpha }$ having similar structures as $\delta_{yy},I_{yy},I_{yu_0}$ but with $y(t)$ replaced by $y^\alpha (t)$.\\
\textbf{Step 2:}
Starting with a stabilizing $K_0^\alpha $, \textit{Solve} for $K^\alpha $ iteratively ($k=0,1,\dots$) once matrices $\delta_{y^\alpha y^\alpha },I_{y^\alpha y^\alpha },I_{y^\alpha u_0^\alpha }$ are constructed and iterative equation can be written for each small learning steps as,
\begin{align}\label{eq:update}
\hspace{-.3 cm} \underbrace{\begin{bmatrix}
\delta_{y^\alpha y^\alpha} & -2I_{y^\alpha y^\alpha}( I_s \otimes K_k^{\alpha T}R^\alpha)  -2I_{y^\alpha u_0^\alpha }(I_s \otimes R^\alpha)
\end{bmatrix}}_{\Theta_k^\alpha } \times \begin{bmatrix}
vec(P_k^\alpha ) \\ vec(K_{k+1}^\alpha)
\end{bmatrix}  
  =\underbrace{-I_{y^\alpha y^\alpha }vec(Q_k^\alpha )}_{\Phi_k^\alpha }.
\end{align}
The stopping criterion for this update is $\norm{P_k^\alpha - P_{k-1}^\alpha} < \gamma_2$, where $\gamma_2$ is a chosen small positive threshold.\\
\textbf{Step 3:} Next $\tilde{u}^\alpha=-K^\alpha y^\alpha $ is applied and $u_0^\alpha$ source is removed.\\
\textbf{End For}
\end{algorithm}
\normalsize
\vspace{-.35 cm}
\subsubsection{Analysis and Stability for the Decentralized design}

In this section we analyze the sub-optimality and stability aspects of the area-wise decentralized  controller learned from Algorithm $2$. The learned controller $K^{\alpha} \in \mathbb{R}$ for all the areas will be perturbed from the controller computed using  $y_d^\alpha$, i.e.,
\begin{align}
P^\alpha=\bar{P}^\alpha + \Delta P^\alpha ,
K^\alpha=\bar{K}^\alpha + \Delta K^\alpha, 
\end{align}
where $\bar{P}^\alpha,\bar{K}^\alpha$ are the optimal solutions if the clusters were decoupled and $y_d^\alpha(t)$ were available for design, and $\Delta P^\alpha, \Delta K^\alpha$ are matrix perturbations. 
The following theorem shows that the matrix perturbations are $O(\epsilon)$ small. \\
\textbf{Theorem 6:}  \textit{Assuming $||y_d^\alpha(t)||$ and $||u_0^\alpha(t)||$ are bounded, the area-wise decentralized solutions satisfy for $\alpha = 1,\dots,r$
\vspace{-.4 cm}
\begin{align}\label{decen_sub}
\hspace{-.4 cm} P^\alpha=\bar{P}^\alpha + O(\epsilon),
K^\alpha=\bar{K}^\alpha + O(\epsilon) ,\; J^\alpha = \bar{J}^\alpha + O(\epsilon).
\vspace{-.35 cm}
\end{align}}
\noindent \textit{Proof:}
This proof directly follows from the analysis performed for Theorem $2$. Here the time-scale separation exists between the decoupled average variable $y_d^\alpha$ and the actual average variable $y^\alpha$. Using Lemma $2$, these variables are $O(\epsilon)$ apart, which leads to \eqref{decen_sub} following the analysis of Theorem 2, and Corollary $1$. \qed 

Next we analyze the closed-loop stability conditions for the block-decentralized design.
\par

\noindent \textbf{Theorem 7:} 
\textit{Assume that the control policy $u^\alpha = -M^\alpha K_k^\alpha y^\alpha$ for area $\alpha$ at the $k^{th}$ iteration is asymptotically stable. Then the control policy at the $(k+1)^{th}$ iteration given by $u^\alpha =- M^\alpha K_{k+1}^\alpha y^\alpha$ is asymptotically stable with $R^\alpha \succ 0$ and $Q^\alpha \succ 0$, if $\epsilon $ is sufficiently small.}\qed 

\noindent \textit{Proof:} The proof is given in Appendix E. 
\vspace{-.4 cm}
\section{Numerical Simulations} 

\subsection{Centralized State Feedback Design}
\begin{figure*}
  \begin{minipage}{.33\linewidth}
    { \includegraphics[width=.9\linewidth, height= 2.3 cm, trim = 4 4 4 4,clip]{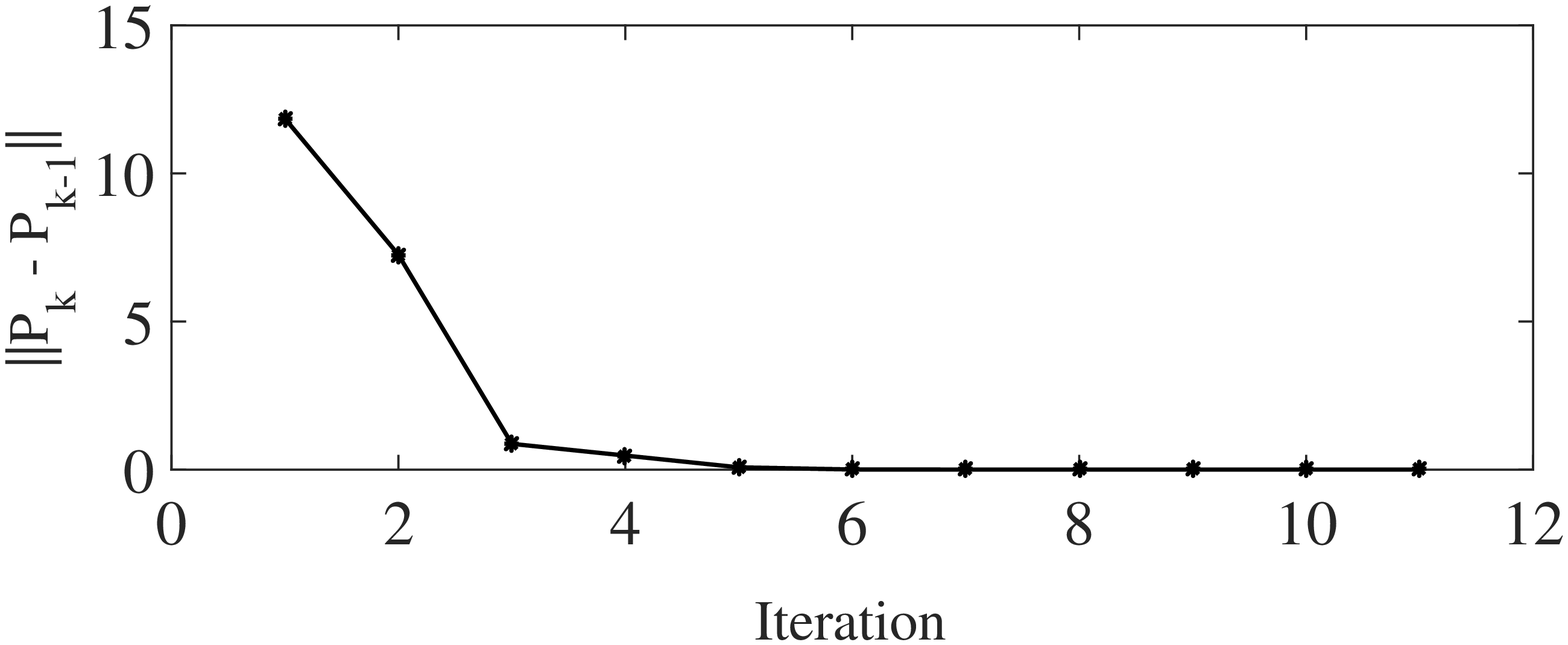}}
    \includegraphics[ width=.9\linewidth, height= 2.3 cm, trim = 4 4 4 4,clip]{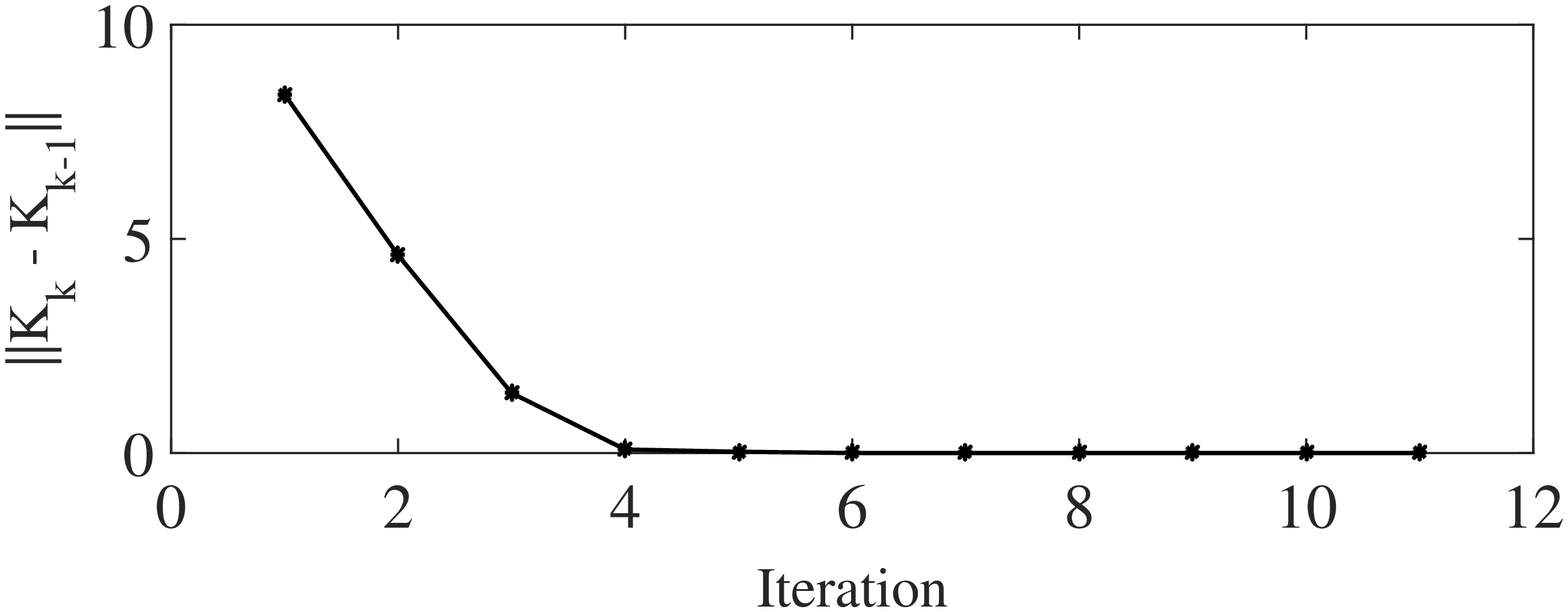}
\caption{\small{Convergence of $P$ \protect\\ and $K$ for the standard \protect\\ SP system \protect\\ }}
\label{fig:SP2}
  \end{minipage}%
  \begin{minipage}{.33\linewidth}
    \includegraphics[width=.9\linewidth, height= 2.3 cm,, trim = 4 4 4 4,clip]{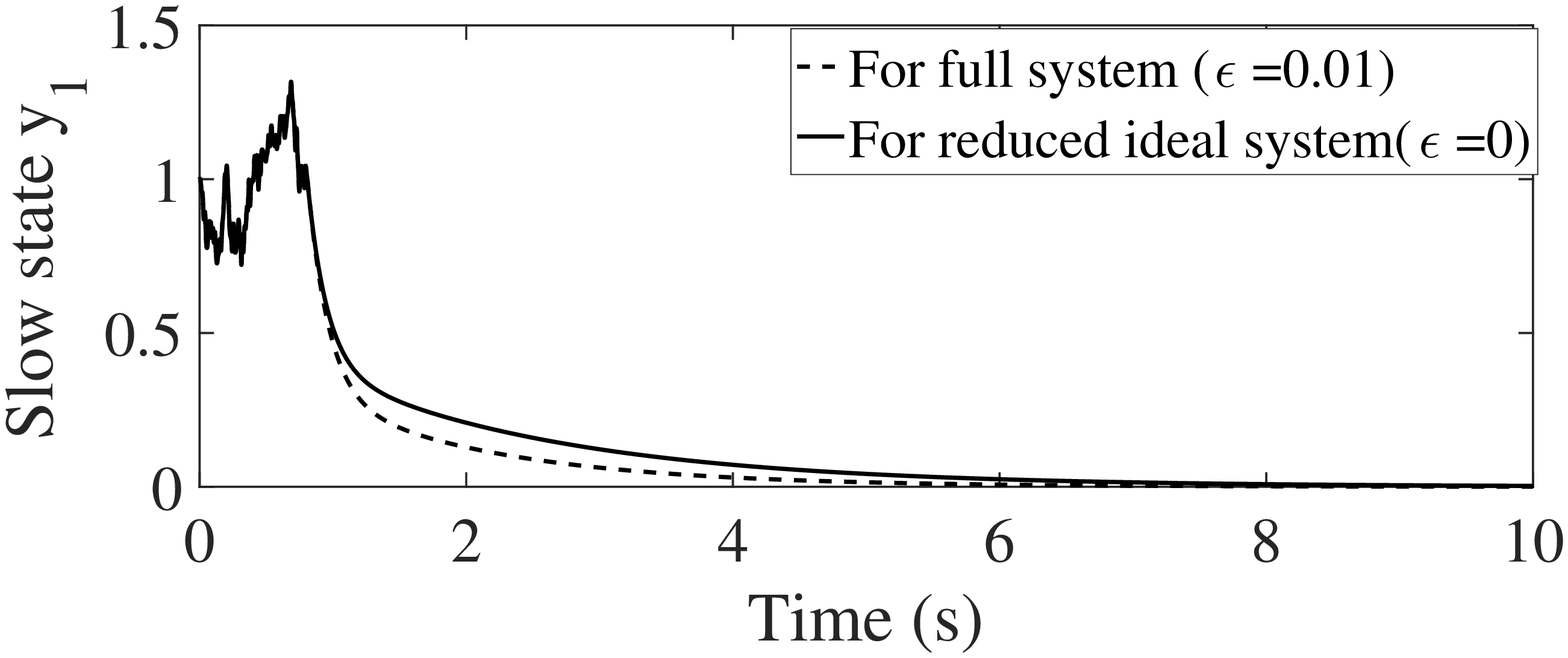}
\includegraphics[width=.9\linewidth, height=2.3 cm,, trim = 4 4 4 4,clip]{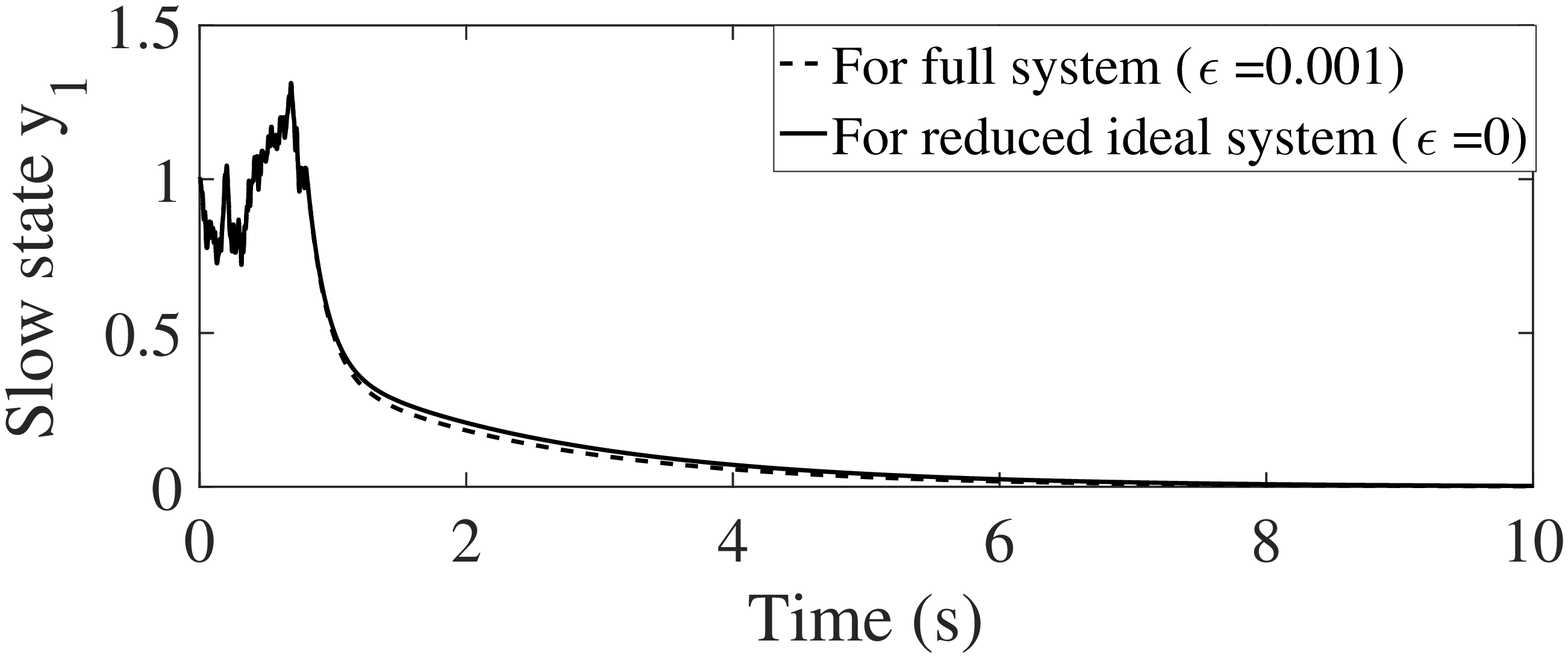}
\caption{\small{Comparison of slow state $1$ with $\epsilon = 0.01,0.001$ and reduced slow subsystem \protect\\}}
\label{fig:SP3}
  \end{minipage}
  \begin{minipage}{.33\linewidth}
    \includegraphics[width=.9\linewidth, height= 2.3 cm,, trim = 4 4 4 4,clip]{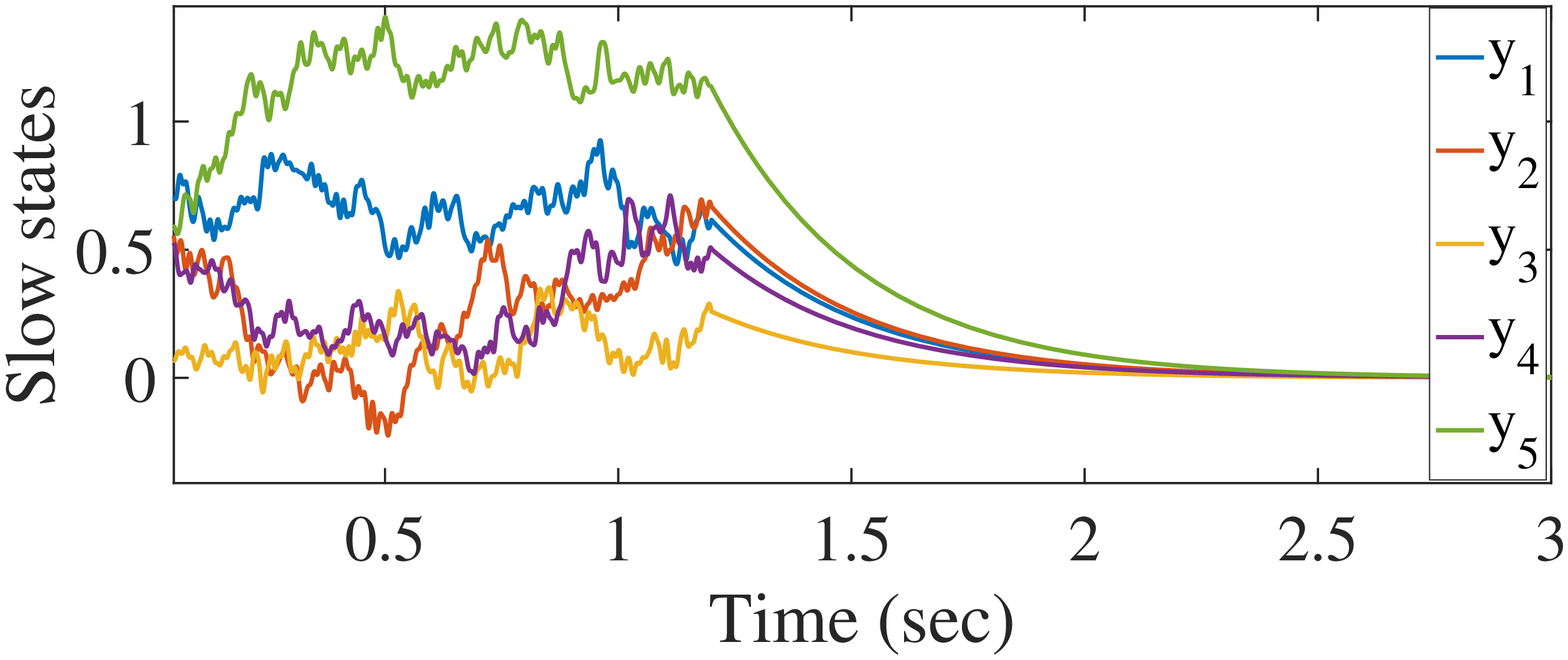}
\includegraphics[width=.9\linewidth, height= 2.3 cm,, trim = 4 4 4 4,clip]{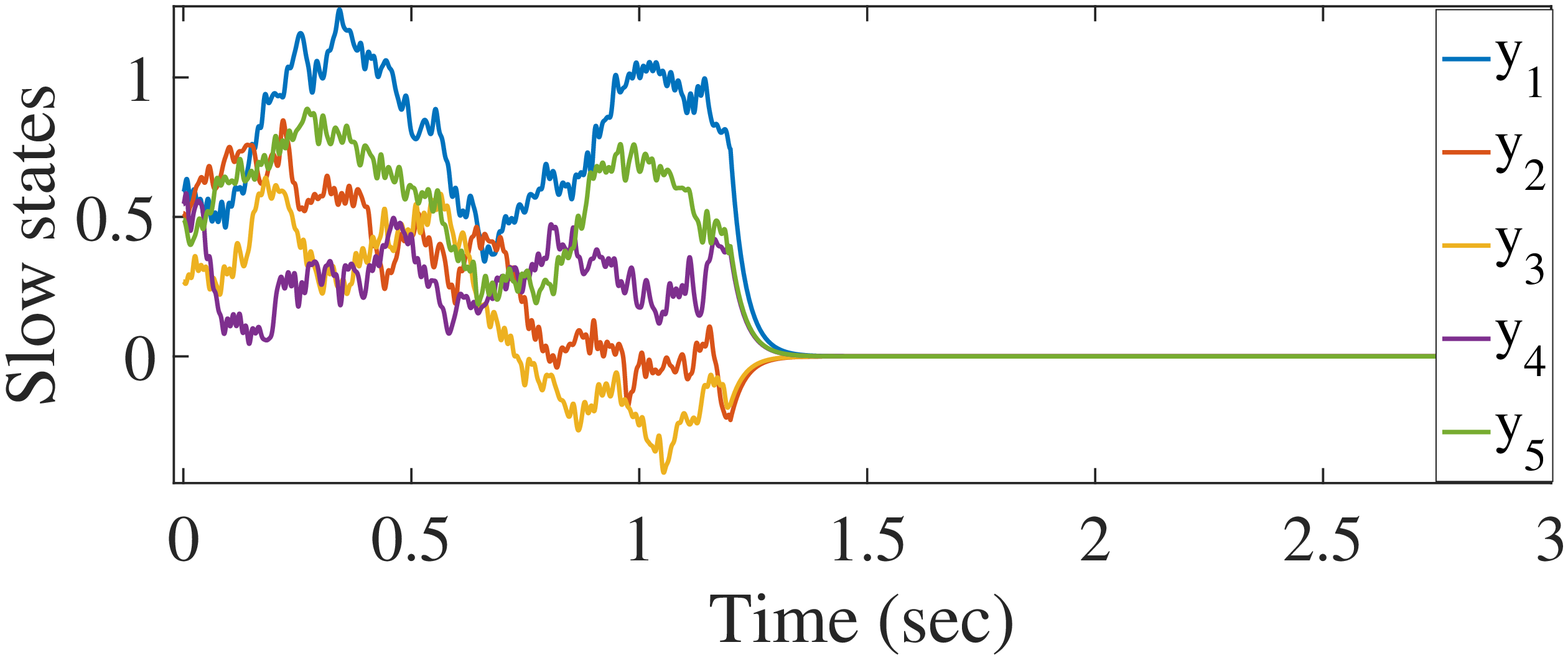}
\caption{\small{Improved dynamics for the clustered network (Top panel - $Q=10I_5$, Bottom panel - $Q=1000I_5$)}}
\label{fig:cluster}
  \end{minipage}
    \bigskip
\begin{minipage}{.33\linewidth}
    { \includegraphics[width=.9\linewidth, height= 2.3 cm, trim = 4 4 4 4,clip]{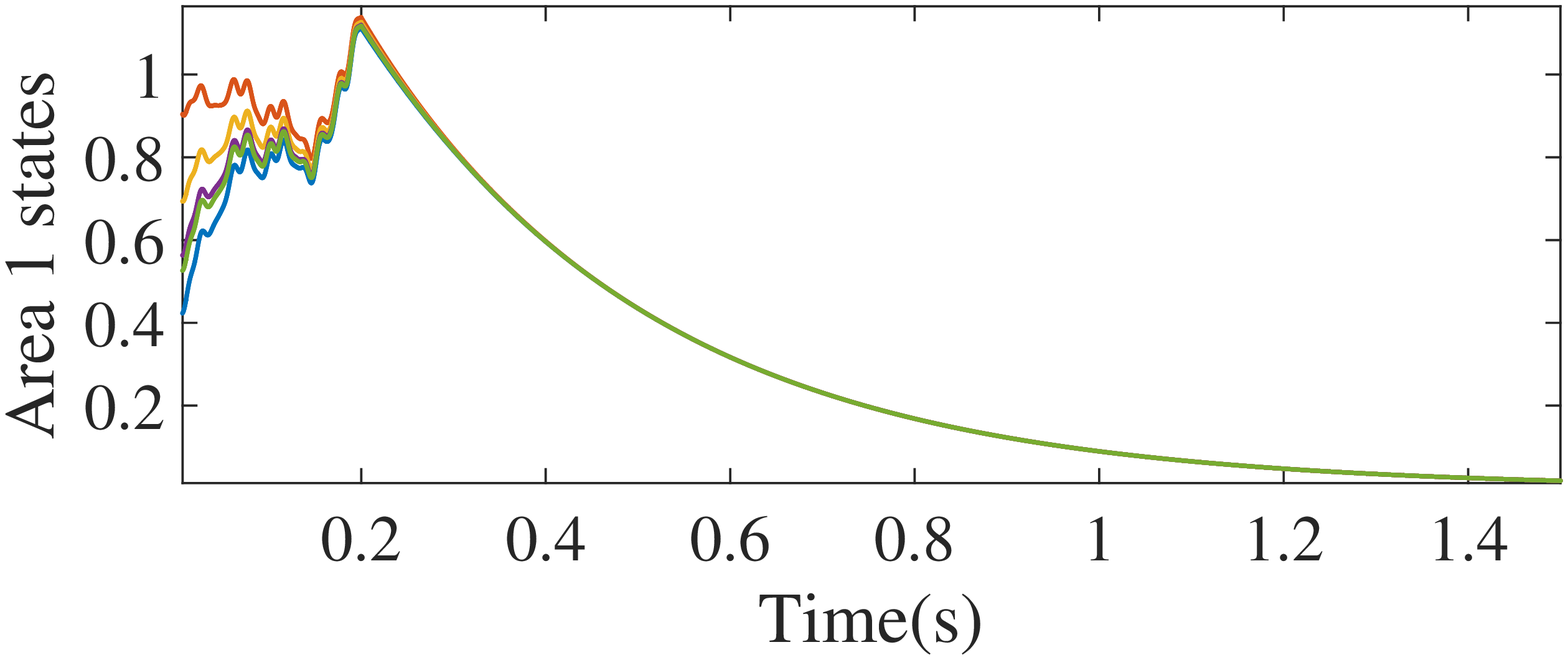}}
    \centering
\caption{\small{ Decentralized\protect\\ design for ideal decoupled clusters\protect\\  \protect\\}}
\label{fig:iso}
  \end{minipage}%
  \begin{minipage}{.33\linewidth}
    \includegraphics[width=.9\linewidth, height= 2.3 cm,, trim = 4 4 4 4,clip]{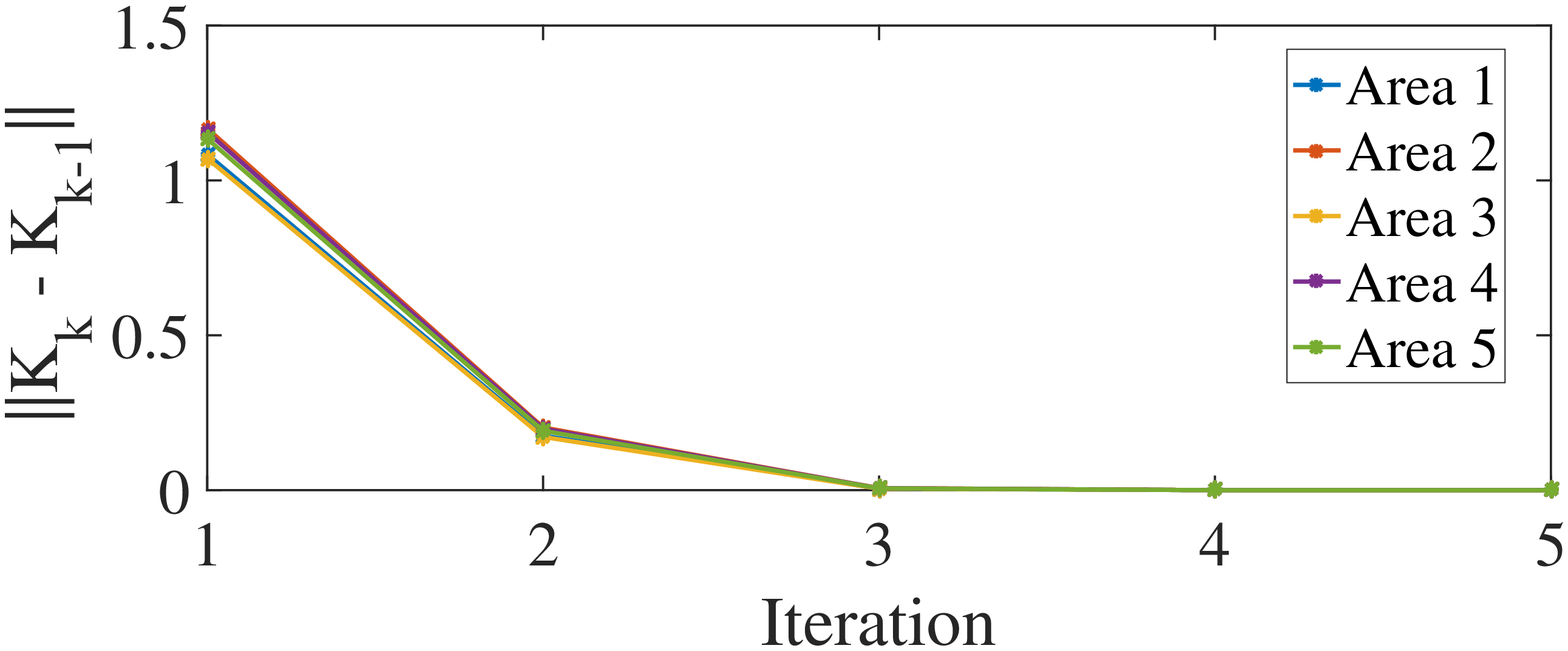}
\caption{\small{Convergence of $K$ and $P$ for the cluster-wise decentralized design \protect\\}}
\label{fig:KP}
  \end{minipage}
  \begin{minipage}{.33\linewidth}
\includegraphics[width=.9\linewidth, height= 2.3 cm,, trim = 4 4 4 4,clip]{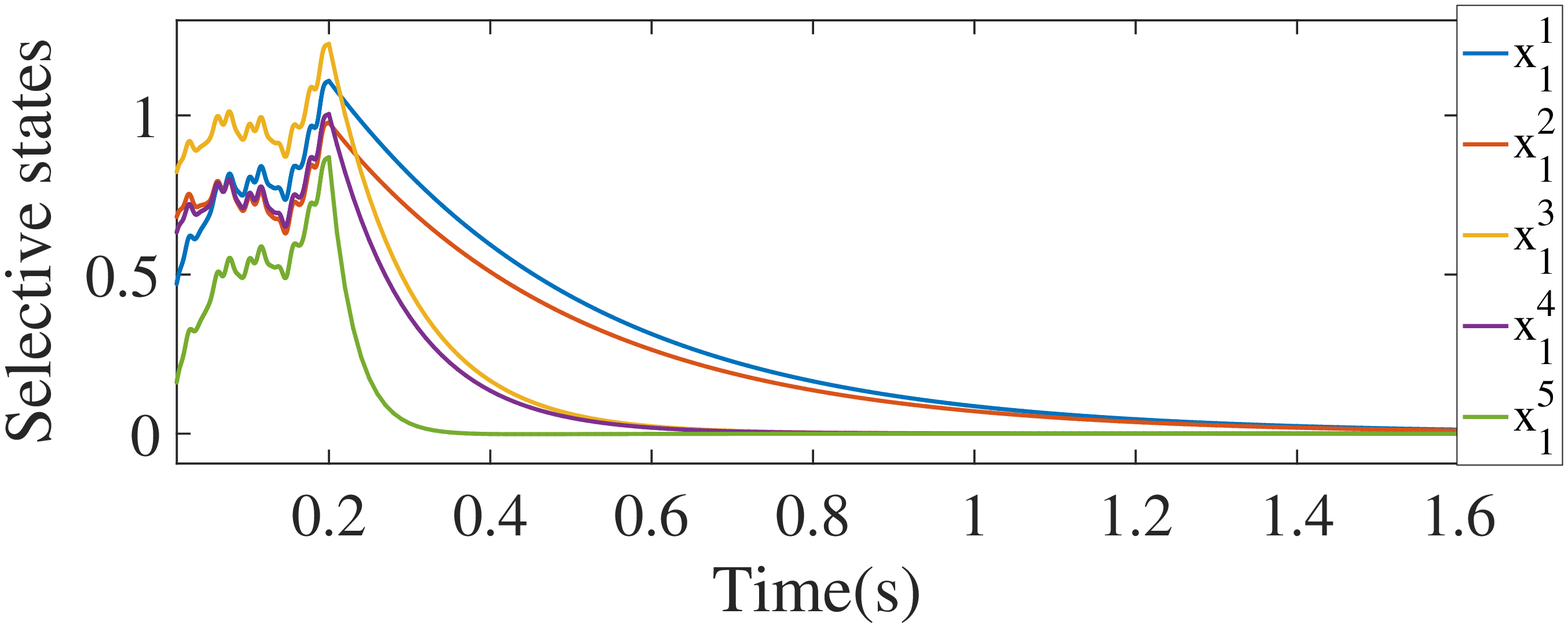}
\caption{\small{Dynamic performance with cluster-wise decentralized design (Top panel - $Q = 10I_5$ for all areas, Bottom panel - varying $Q$ }}
\label{fig:decen}
  \end{minipage}
  \vspace{-.5 cm}
\end{figure*}

A singularly perturbed system in the form of~\eqref{eq:SP} is considered with two fast and two slow states. We choose $\epsilon =0.01$, $Q=10I_2, R=I$, the initial conditions as $[1,\,2,\,1,\,0]$, and the learning time-step as $0.01$ seconds. The model matrices are taken from \citet{chowslowfast} as
\begin{align*}
& A_{11}=\begin{bmatrix}0 & 0.4 \\ 0 & 0 \end{bmatrix},\; A_{12}=\begin{bmatrix}0 & 0 \\ 0.345 & 0 \end{bmatrix},\;A_{21}=\begin{bmatrix}0 & -0.524 \\ 0 & 0 \end{bmatrix}\\
& A_{22}=\begin{bmatrix}-0.465 & 0.262 \\ 0 & -1 \end{bmatrix},\; B_1 = B_2 =\begin{bmatrix}1\\ 1 \end{bmatrix}.
\end{align*}
\normalsize
The system is persistently excited by exploration noise following \citet{jiang_book}. The control gain is learned as $K = [3.80 \;\;  1.38]$, producing a closed-loop objective $J=7.72$ units. The convergence plots for $P$ and $K$ are shown in Fig.~\ref{fig:SP2}.
We next compare the closed-loop responses learned by ADP for the ideal reduced slow system ($\epsilon = 0$) versus the full-order system ($\epsilon \neq 0$) in Fig.~\ref{fig:SP3}. For the ideal slow system, the following controller is learned: $\bar{K} = [3.1623 \;\;  1.9962], \bar{J}= 7.2950$ units. The top panel of Fig.~\ref{fig:SP3} shows this comparison for $\epsilon=0.01$, while the bottom panel shows this for $\epsilon=0.001$. It can be seen that the responses of the ideal and non-ideal reduced-dimensional systems get closer to each other over time as $\epsilon$ decreases.\par
We next consider a clustered multi-agent network with $25$ agents, divided into $5$ clusters. Each agent has a scalar state with $F=0$. Therefore the network has $4$ slow eigenvalues, one zero eigenvalue and the rest are the fast eigenvalues. The slow eigenvalues are $-0.128,-0.195,-0.196,$ and $-0.2638$. The control architecture is shown in Fig.~\ref{arch_central}. Each cluster is assumed to have a local coordinator that averages the states from inside the cluster, and transmits the average state to a central controller, which learns the reduced-dimensional control input $\tilde{u}(t) \in \mathbb R^5$ and subsequently back-projects it to individual agents.
\begin{figure}[H]
\centering
\includegraphics[width=.7\linewidth, trim = 4 4 4 4,clip]{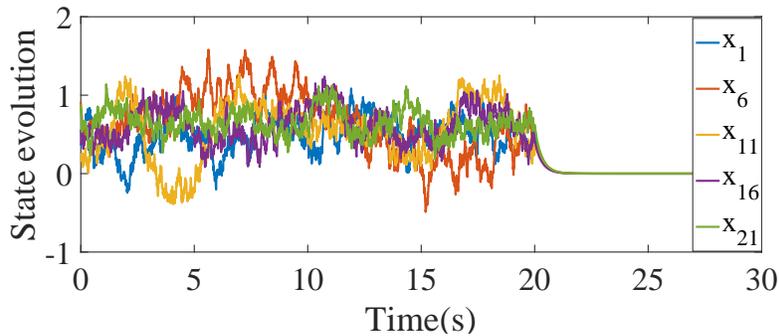}
\caption{\small{Learned controller with full-state feedback}}
\label{fig:full}
\vspace{-.3 cm}
\end{figure} 
Fig.~\ref{fig:full} shows the learning of the full-dimensional optimal LQR controller. It takes at least $18.75$ seconds to learn $K \in \mathbb{R}^{25\times 25}$. The exploration signal here is a sum of sinusoidal signals with different frequencies. With $r=5$, the reduced-dimensional controller, on the other hand, requires only $r^2 + 2r^2 = 75$ samples for learning. It dominantly affects the slow poles, and with $Q=10I_5$, the closed-loop slow poles are placed at $-3.14,-3.18,-3.17,-3.15,$ and $-3.16$. {Dynamic performance is improved with increase in the weights of $Q$ as shown in Fig.~\ref{fig:cluster}.  }
A comparison between the full and the reduced-dimensional design in terms of minimum learning and CPU run times is given in Table $1$.
\begin{table}[H]
\centering
\caption{\normalsize{Reduction in learning and CPU run times for the slow state feedback-based design with $25$ agents}}
\label{label1}
\begin{tabular}{l|l|l|}
\cline{2-3}
                                                                                                                & \begin{tabular}[c]{@{}l@{}}\scriptsize{Ideal min. learning} \\ \scriptsize{time (T=0.01 s)}\end{tabular} & \begin{tabular}[c]{@{}l@{}}\scriptsize{CPU run} \\ \scriptsize{times}\end{tabular} \\ \hline
\multicolumn{1}{|l|}{\begin{tabular}[c]{@{}l@{}} \scriptsize{Full-state feedback}\end{tabular}}   & \scriptsize{18.75 s}                                                                        & \scriptsize{72.19 s}                                                                  \\ \hline
\multicolumn{1}{|l|}{\begin{tabular}[c]{@{}l@{}}\scriptsize{Reduced-dim state feedback}\end{tabular}} & \scriptsize{0.75 s}                                                                         & \scriptsize{1.34 s}                                                                   \\ \hline
\end{tabular}
\end{table}
\vspace{-.4 cm}
\subsection{Cluster-wise decentralized state feedback design} 
Considering the same multi-agent example, we first perform the ADP-based learning of the controller when the clusters are fully decoupled (i.e., the ideal decentralized  scenario). Each area is equipped with an aggregator. 
Note that the average of all the cluster states represents the decoupled slow state $y_d^\alpha$ for cluster $\alpha$. The state evolution of two representative areas are shown in Fig.~\ref{fig:iso}. We consider similar coupling strengths between the agents inside all the clusters  with $Q=10,\, R=1$ but with different initial conditions. The computed scalar control gain for each area is $K=3.1623$, and the corresponding objective values are $\bar{J}^1 = 1.317,\,\bar{J}^2 = 0.745,\, \bar{J}^3 = 1.765,\, \bar{J}^4 = 0.8451$ and $\bar{J}^5 = 0.5244$. 
\par 
Thereafter, the decentralized ADP computation is performed on the actual system following Algorithm $2$. The average states from each cluster  is used as the feedback signal for the ADP computation block as shown in Fig.~\ref{arch_decen}.  Fig.~\ref{fig:KP} shows the fast convergence of the ADP iterations.  
With $Q=10,R=1$ for all the areas, the cluster-wise decentralized control gains are computed as $K^1=3.139,\,K^2=3.195, \, K^3=3.130, \, K^4 = 3.187, \, K^5=3.173,$ with the objective values as $J^1=1.308,\, J^2= 0.754,\, J^3= 1.7478, \, J^4=0.8524$ and $J^5=0.5261$. In Fig.~\ref{fig:decen}, we can see that with the increasing value of $Q^\alpha,\alpha = 1,\dots,5$, the dynamic performance of the agent states increases. The dynamic performance of different cluster states can be controlled independently using different $Q$ for the different areas. The learning time is also decreased because of the reduced number of feedback variables. The exploration is performed for only $0.2$ seconds.
\vspace{-.3 cm}
\subsection{Output feedback RL (OFRL) design}
\begin{figure}[t]

\begin{minipage}{.33\linewidth}
    \includegraphics[width=.9\linewidth, height= 2.3 cm,, trim = 4 4 4 4,clip]{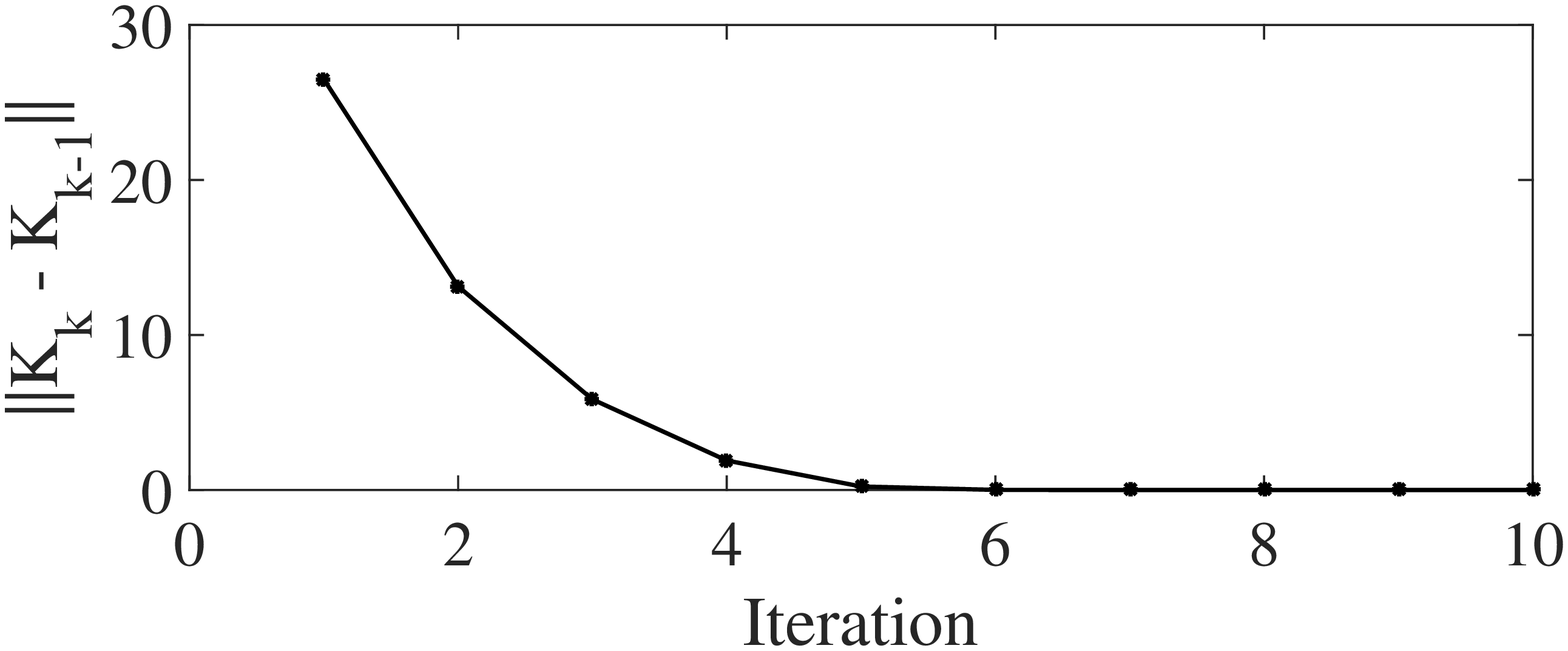}
\includegraphics[width=.9\linewidth, height= 2.3 cm, trim = 4 4 4 4,clip]{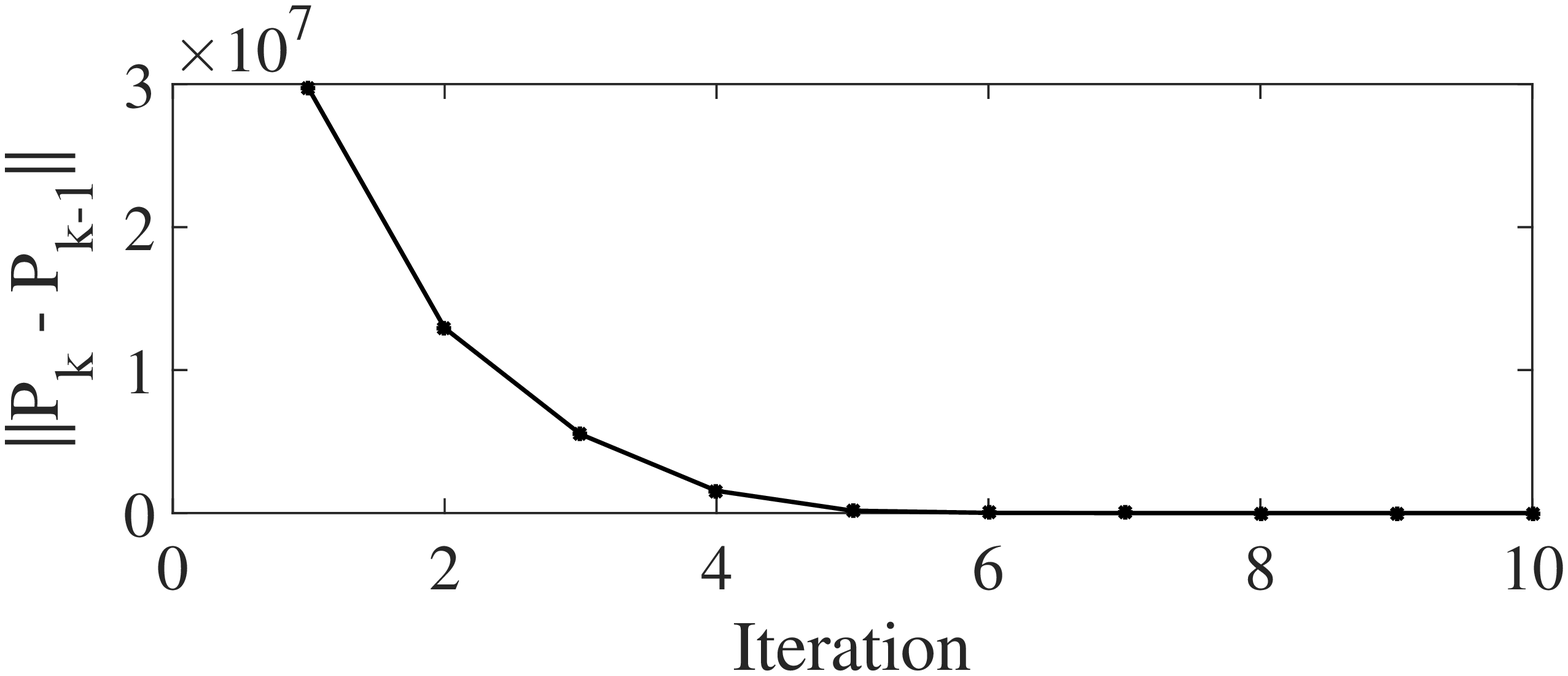}
\caption{\small{Convergence of K and P for the standard SP system (OFRL)}}
\label{fig:iteration}
\end{minipage}
\begin{minipage}{.33\linewidth}
\includegraphics[width=.9\linewidth, height= 2.3 cm,, trim = 4 4 4 4,clip]{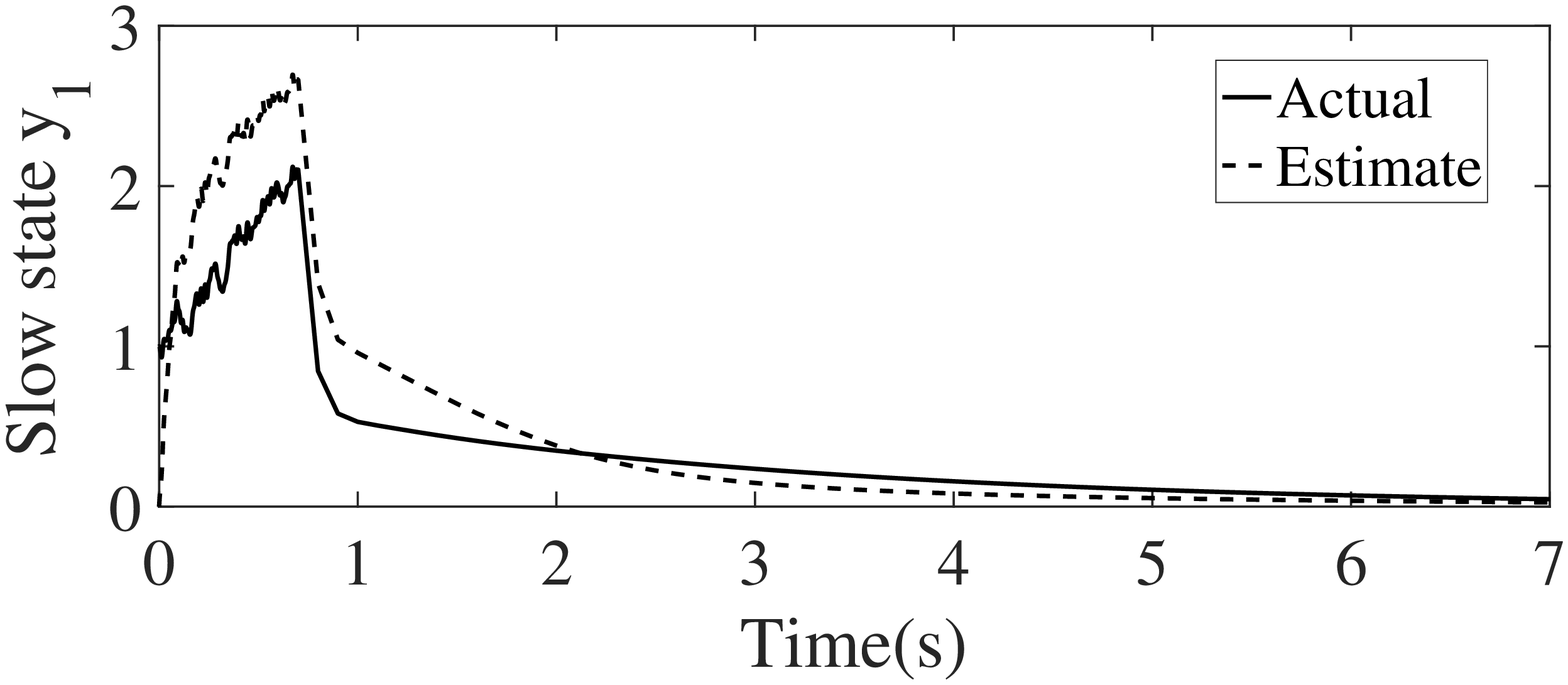}
\includegraphics[width=.9\linewidth, height= 2.3 cm,, trim = 4 4 4 4,clip]{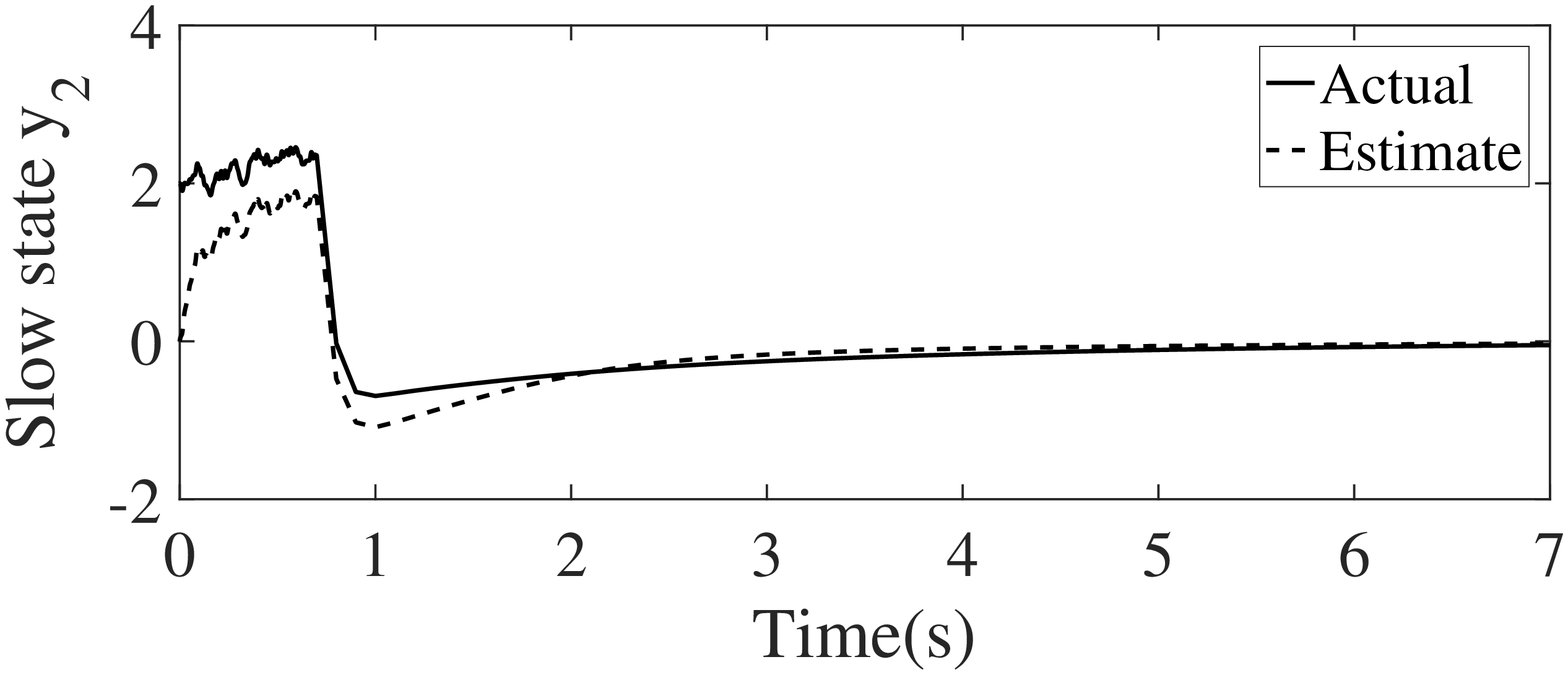}
\caption{\small{Slow states for the \protect\\ standard SP system (OFRL)}}
\label{fig:spslow}
  \end{minipage}%
  \begin{minipage}{.33\linewidth}
   \includegraphics[width=.9\linewidth, height= 2.3 cm,, trim = 4 4 4 4,clip]{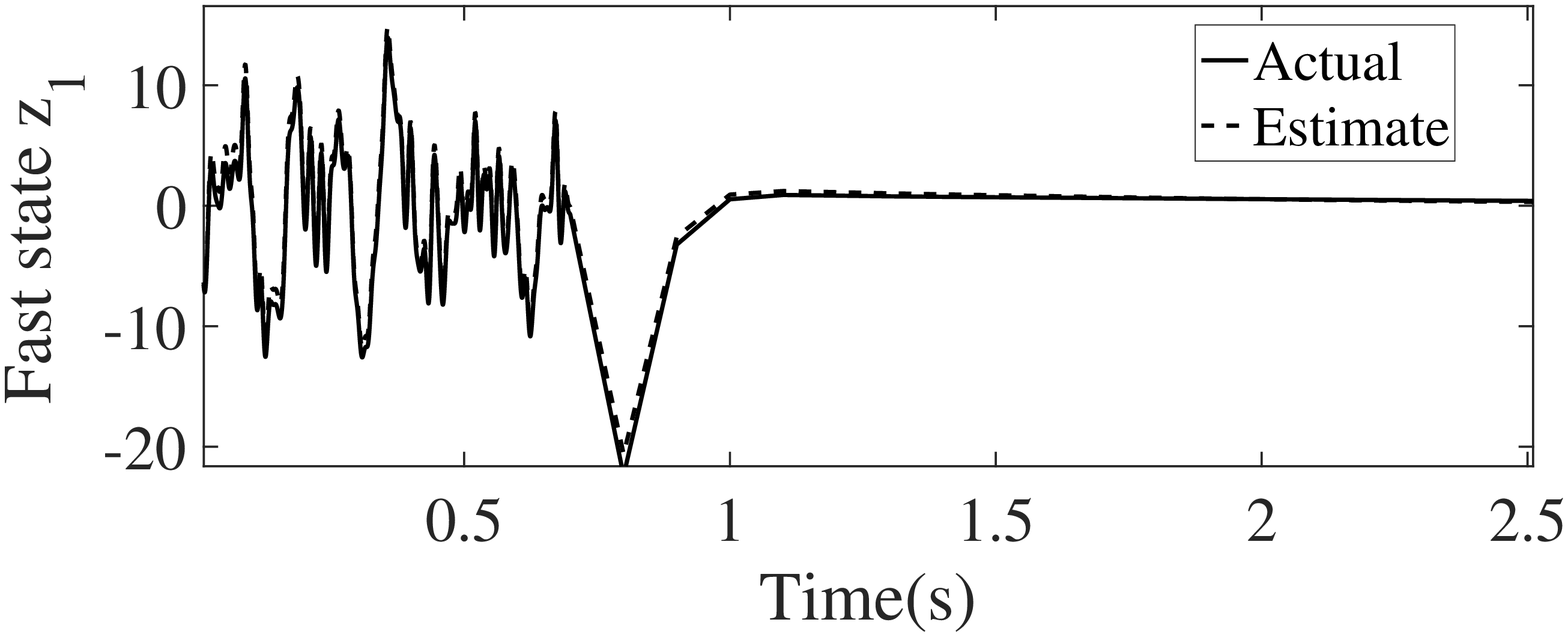}
\includegraphics[width=.9\linewidth, height= 2.3 cm,, trim = 4 4 4 4,clip]{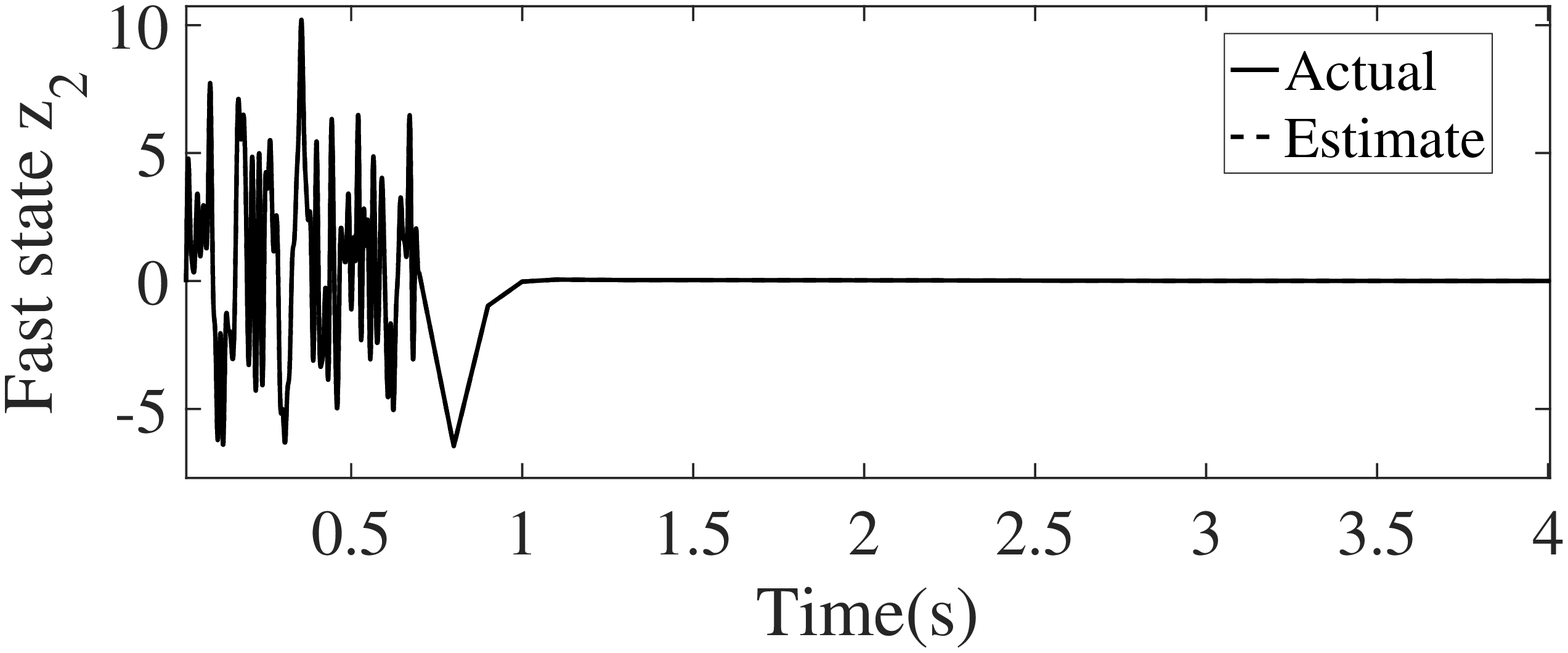}
\caption{\small{Fast state trajectories for the standard SP system (OFRL)}}
\label{fig:spfast}
\end{minipage}
\bigskip
\begin{minipage}{.33\linewidth}
\includegraphics[width=.9\linewidth, height= 2.3 cm,, trim = 4 4 4 4,clip]{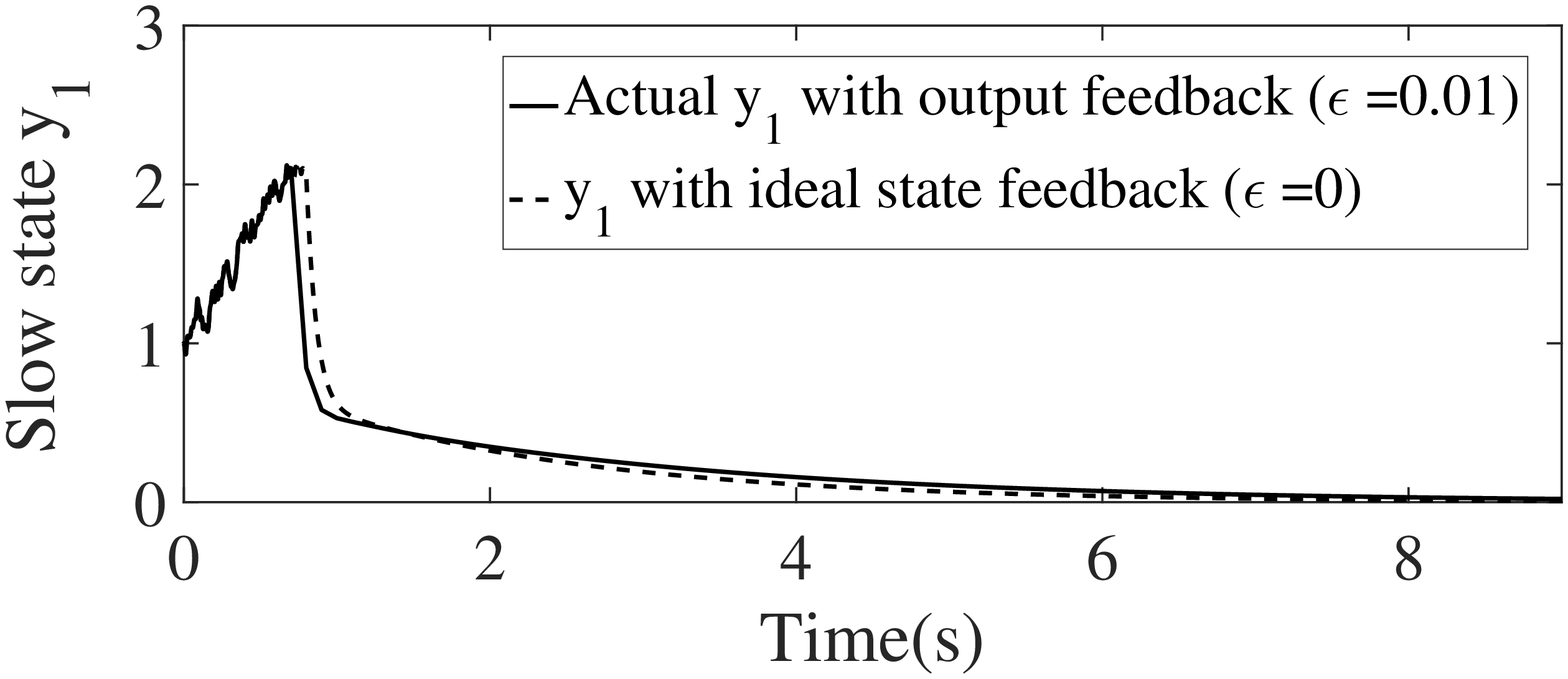}
\includegraphics[width=.9\linewidth, height= 2.3 cm,, trim = 4 4 4 4,clip]{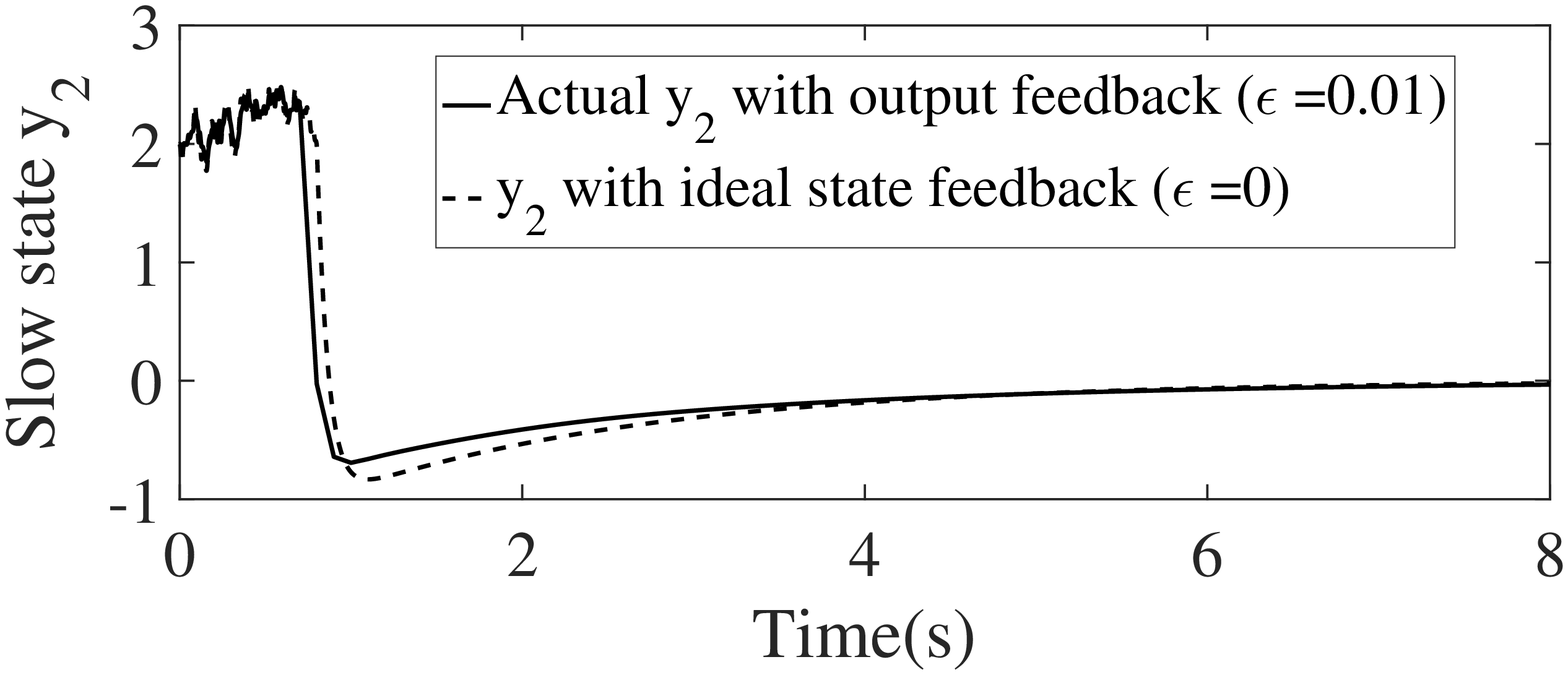}
\caption{\small{Comparison with \protect\\state feedback for the $\epsilon =0$ \protect\\ system (OFRL) }}
\label{fig:spcomp}
  \end{minipage}
  \begin{minipage}{.33\linewidth}
   \includegraphics[width=.9\linewidth, height= 2.3 cm,, trim = 4 4 4 4,clip]{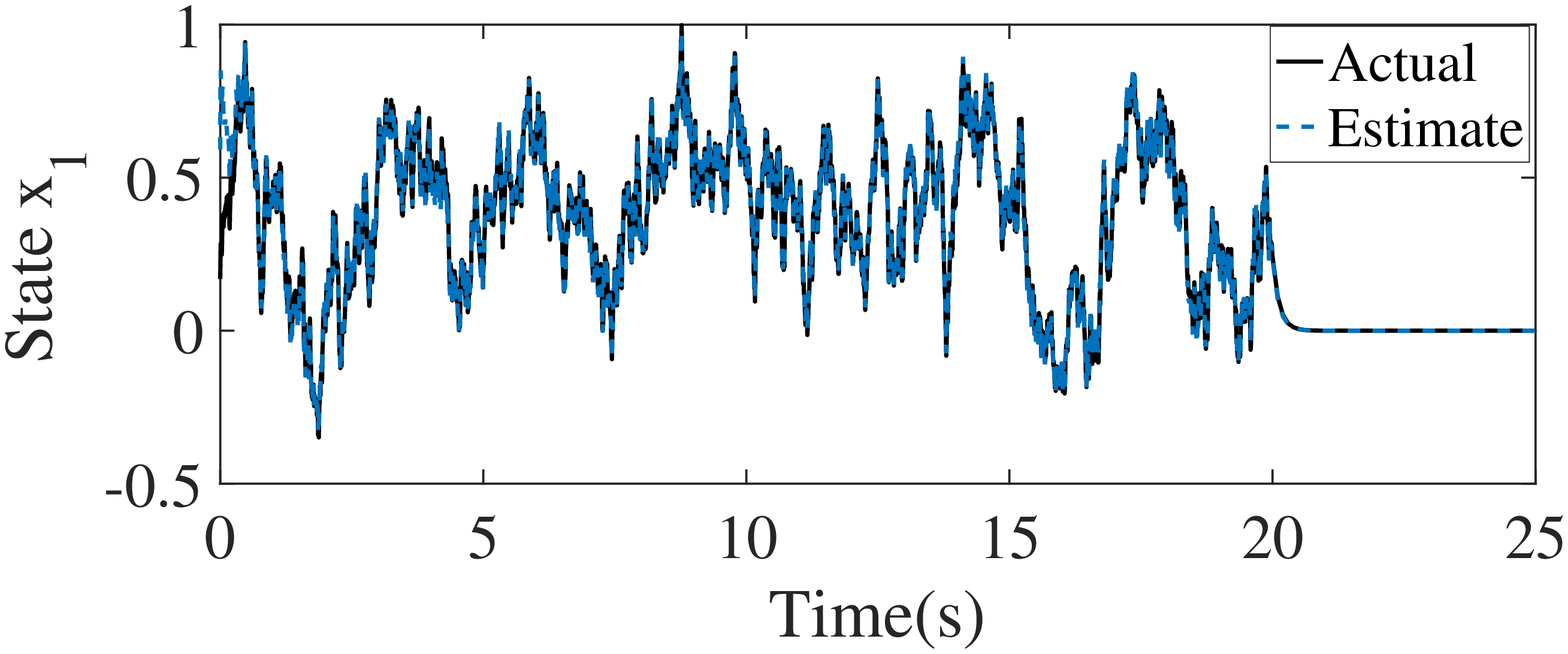}
\includegraphics[width=.9\linewidth, height= 2.3 cm,, trim = 4 4 4 4,clip]{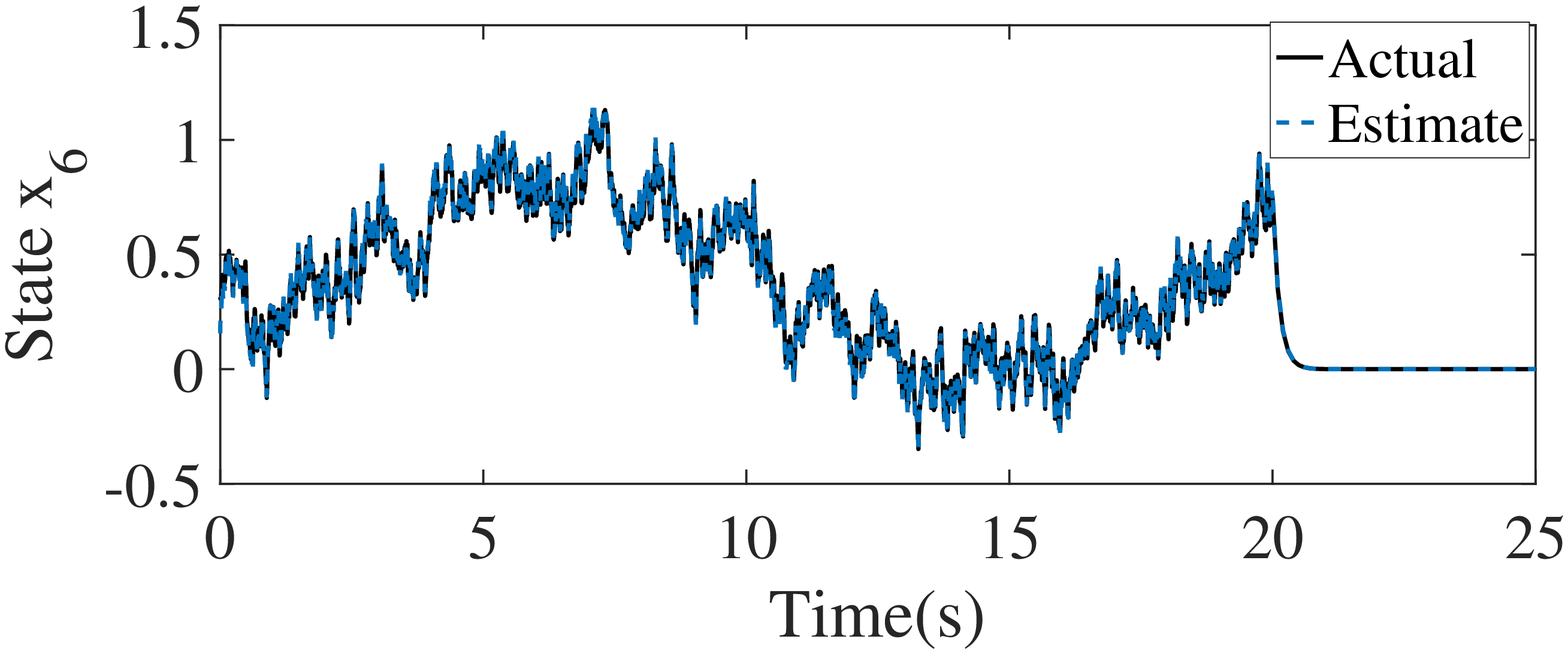}
\caption{\small{Learning with full \protect\\ state estimates for the \protect\\clustered network}}
\label{fig:clusterfull}
\end{minipage}
\begin{minipage}{.33\linewidth}
\includegraphics[width=.92\linewidth, height= 2.3 cm,, trim = 4 4 4 4,clip]{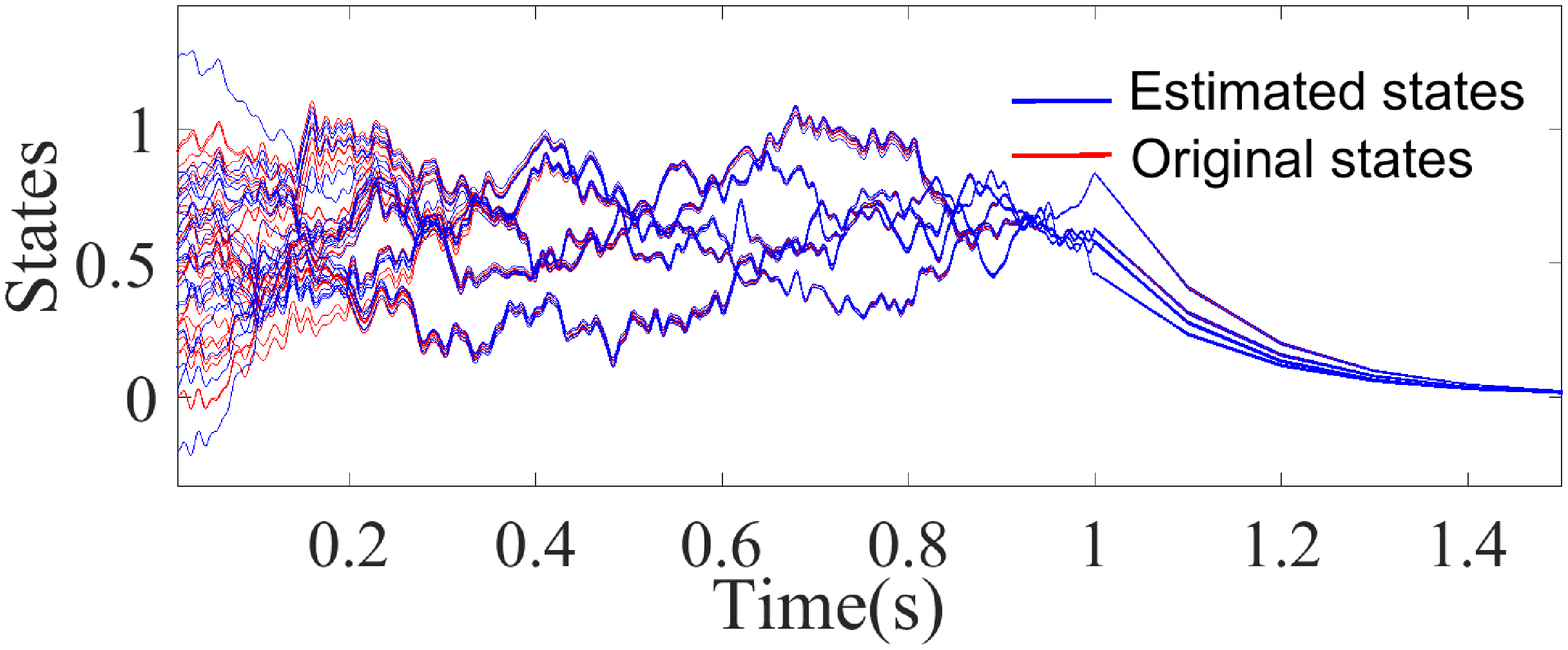}
\includegraphics[width=.9\linewidth, height= 2.3 cm,, trim = 4 4 4 4,clip]{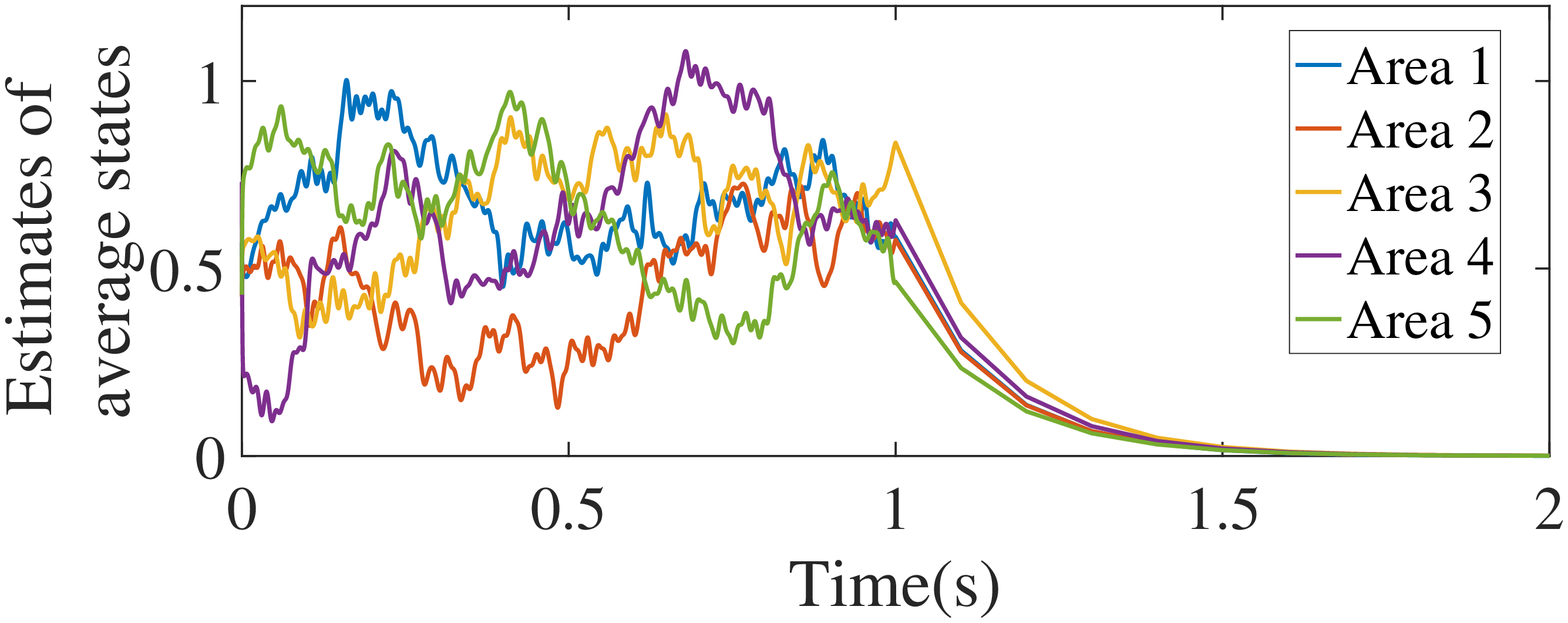}
\caption{\small{Learning with slow state estimates for the clustered network}}
\label{fig:clusterred}
  \end{minipage}
  \vspace{-1 cm}
\end{figure}

We first consider the singularly perturbed system as in Section $7.1$ with $\epsilon =0.01$, initial condition $[1,\,2,\,1,\,0]$. We consider $C=[1,1,0,0;0,0,1,1]$. The learning time step is $0.01$ seconds. Data is gathered for $0.7$ s with the system being persistently excited with exploration noise. Fig.~\ref{fig:iteration} shows the convergence of $P$ and $K$ during the ADP-based computations using the estimated states. Fig. \ref{fig:spslow} and Fig.~\ref{fig:spfast} show the actual versus estimated state trajectories using the NN observer. 
For the design of the NN observer, the Hurwitz matrix $\hat{A}$ is considered to be of SP structure but different than the original state matrix. We can see from Figs. \ref{fig:spslow}-\ref{fig:spfast} that the estimation error is small, and the ADP controller using these estimates maintains closed-loop stability. Also, Fig.~\ref{fig:spcomp} compares the output feedback control responses with the ideal ($\epsilon =0$) state feedback responses. 
\par
We next consider the $5$-cluster, $25$-agent clustered consensus network.
We consider a slightly different set of couplings with similar structure as considered for the state feedback design. The slow eigenvalues are $-0.127,-0.192,-0.191$ and $-0.258$. 
For the estimator design, the Hurwitz matrix $\hat{A}$ is taken to be of similar structure as $A$ but the coupling between the agents in a same cluster is $20 \% $ off from the original, while the inter-cluster strengths are $50 \%$ off from the original. For the full-order system, Fig.~\ref{fig:clusterfull} shows few examples of the state estimation, where the learning takes approximately $20$ s. In the reduced-dimensional design, using the NN observer estimates  the aggregator generates the average states for each cluster. These average states and inputs are used for the reduced-dimensional ADP iterations. Fig.~\ref{fig:clusterred} shows that the reduced-dimensional design using the NN observer requires approximately $1$ s of exploration. The comparison of learning and CPU run-times between the full-dimensional observer-based design and the reduced-dimensional  observer-based design is presented in Table \ref{table2}.
\begin{table}[H]
\centering
\caption{\normalsize{Reduction in learning and CPU run times for the output feedback based reduced order design with $25$ agents}}
\label{table2}
\begin{tabular}{l|l|l|}
\cline{2-3}
                                                                                                              & \begin{tabular}[c]{@{}l@{}} \scriptsize{Ideal min. learning}\\ \scriptsize{time (T=0.01 s)}\end{tabular} & \begin{tabular}[c]{@{}l@{}}\scriptsize{CPU run}\\ \scriptsize{ times} \end{tabular} \\ \hline
\multicolumn{1}{|l|}{\begin{tabular}[c]{@{}l@{}} \scriptsize{Full-dim output feedback} \end{tabular}}   & \scriptsize{18.75 s}                                                                        & \scriptsize{298 s}                                                                        \\ \hline
\multicolumn{1}{|l|}{\begin{tabular}[c]{@{}l@{}} \scriptsize{Reduced-dim output feedback}\end{tabular}} & \scriptsize{0.75 s}                                                                         & \scriptsize{13.82 s}                                                                      \\ \hline
\end{tabular}
\end{table}
 \vspace{-.6 cm}
\section{Conclusion}
The paper presented RL based optimal control designs incorporating ideas from model reduction following from time-scale separation properties in LTI systems. Both state feedback and output feedback RL designs are reported. The designs are extended to clustered multi-agent networks for which an additional cluster-wise block-decentralized RL control is also discussed. Sub-optimality and stability analyses for each design are performed using SP approximation theorems. For the state feedback designs only the SP approximation error affects the sub-optimality, whereas for the output feedback designs the state estimation error adds to it. Results are validated using multiple simulation case studies.

\bibliography{ref_cdc}  

\begin{thebibliography}{}

\bibitem[\protect\astroncite{Abdollahi et~al.}{2006}]{obs}
Abdollahi, F., Talebi, H.~A., and Patel, R.~V. (2006).
\newblock A stable neural network-based observer with application to
  flexible-joint manipulators.
\newblock {\em IEEE Transactions on Neural Networks}, 17(1):118--129.

\bibitem[\protect\astroncite{Chow and Kokotovic}{1976}]{chowslowfast}
Chow, J. and Kokotovic, P. (1976).
\newblock A decomposition of near-optimum regulators for systems with slow and
  fast modes.
\newblock {\em IEEE Trans. on Automatic Control}, 21(5):701--705.

\bibitem[\protect\astroncite{Chow and Kokotovic}{1985}]{chow1985}
Chow, J. and Kokotovic, P. (1985).
\newblock Time scale modeling of sparse dynamic networks.
\newblock {\em IEEE Trans. on Automatic Control}, 30(8):714--722.

\bibitem[\protect\astroncite{Jiang and Jiang}{2012}]{jiang1}
Jiang, Y. and Jiang, Z.-P. (2012).
\newblock Computational adaptive optimal control for continuous-time linear
  systems with completely unknown dynamics.
\newblock {\em Automatica}, 48:2699--2704.

\bibitem[\protect\astroncite{Jiang and Jiang}{2017}]{jiang_book}
Jiang, Y. and Jiang, Z.-P. (2017).
\newblock {\em Robust Adaptive Dynamic Programming}.
\newblock Wiley-IEEE press.

\bibitem[\protect\astroncite{Khalil}{2002}]{khalil}
Khalil, H. (2002).
\newblock {\em Nonlinear Systems}.
\newblock Prentice-Hall, New York.

\bibitem[\protect\astroncite{Khalil and Kokotovic}{1978}]{Khalilcontrol}
Khalil, H. and Kokotovic, P. (1978).
\newblock Control strategies for decision makers using different models of the
  same system.
\newblock {\em IEEE Trans. on Automatic Control}, 23(2):289--298.

\bibitem[\protect\astroncite{Kleinman}{1968}]{kleinman}
Kleinman, D. (1968).
\newblock On an iterative technique for riccati equation computations.
\newblock {\em IEEE Trans. on Automatic Control}, 13(1):114--115.

\bibitem[\protect\astroncite{Kokotovic et~al.}{1976}]{SPreduction}
Kokotovic, P., O'malley, R., and Sannuti, P. (1976).
\newblock Singular perturbations and order reduction in control theory: An
  overview.
\newblock {\em Automatica}, 12:123--132.

\bibitem[\protect\astroncite{{Lewis} and {Vamvoudakis}}{2011}]{obs_discrete}
{Lewis}, F.~L. and {Vamvoudakis}, K.~G. (2011).
\newblock Reinforcement learning for partially observable dynamic processes:
  Adaptive dynamic programming using measured output data.
\newblock {\em IEEE Transactions on Systems, Man, and Cybernetics, Part B
  (Cybernetics)}, 41(1):14--25.

\bibitem[\protect\astroncite{{Lewis} and {Vrabie}}{2009}]{lewis_mag}
{Lewis}, F.~L. and {Vrabie}, D. (2009).
\newblock Reinforcement learning and adaptive dynamic programming for feedback
  control.
\newblock {\em IEEE Circuits and Systems Magazine}, 9(3):32--50.

\bibitem[\protect\astroncite{Liu and Wei}{2014}]{nonlin1}
Liu, D. and Wei, Q. (2014).
\newblock Policy iteration adaptive dynamic programming algorithm for
  discrete-time nonlinear systems.
\newblock {\em IEEE Transactions on Neural Networks and Learning Systems},
  25(3):621--634.

\bibitem[\protect\astroncite{Mukherjee et~al.}{a}]{sayak_acc}
Mukherjee, S., Bai, H., and Chakrabortty, A.
\newblock Block-decentralized model-free reinforcement learning control of two
  time-scale networks.
\newblock In {\em American Control Conference 2019}, Philadelphia, PA, USA.

\bibitem[\protect\astroncite{Mukherjee et~al.}{b}]{sayak_cdc}
Mukherjee, S., Bai, H., and Chakrabortty, A.
\newblock On model-free reinforcement learning of reduced-order optimal control
  for singularly perturbed systems.
\newblock In {\em IEEE Conference on Decision and Conrol 2018}, Miami, FL, USA.

\bibitem[\protect\astroncite{Sutton and Barto}{1998}]{barto}
Sutton, R. and Barto, A. (1998).
\newblock {\em Reinforcement learning - An introduction}.
\newblock MIT press, Cambridge, 1998.

\bibitem[\protect\astroncite{Vamvoudakis}{2017}]{V17}
Vamvoudakis, K. (2017).
\newblock Q-learning for continuous-time linear systems: A model-free infinite
  horizon optimal control approach.
\newblock {\em Systems and Control Letters}, 100:14--20.

\bibitem[\protect\astroncite{Vrabie et~al.}{2009}]{vrabie1}
Vrabie, D., Pastravanu, O., Abu-Khalaf, M., and Lewis, F. (2009).
\newblock Adaptive optimal control for continuous-time linear systems based on
  policy iteration.
\newblock {\em Automatica}, 45:477--484.

\bibitem[\protect\astroncite{Wu and Luo}{2012}]{nonlin2}
Wu, H. and Luo, B. (2012).
\newblock Neural network based online simultaneous policy update algorithm for
  solving the {HJI} equation in nonlinear {${H}_{\infty}$} control.
\newblock {\em IEEE Transactions on Neural Networks and Learning Systems},
  23(12):1884--1895.

\end{thebibliography}
\vspace{-.3 cm}
\normalsize

\appendix

\section{Proof of Proposition 1:}
From Lemma $1$ we can write that,
\begin{align}
y_s - c1_r \leq \hat{y} \leq y_s + c1_r.
\end{align}
Here $\leq$ denotes element-wise inequality between vectors. $1_r$ is a $r$-dim vector of all ones. Now using the upper bound of $\hat{y}$ we can introduce slack variable $\Delta(t)$ (with bounded norm) such that $\hat{y} = y_s + c1_r - \Delta$. Then one can get
\begin{align}
\hat{y} \otimes \hat{y}  & = ( y_s + c1_r- \Delta) \otimes (y_s + c1_r- \Delta),\\
& = y_s \otimes y_s + y_s \otimes (c1_r- \Delta) + (c1_r- \Delta) \otimes y_s \nonumber \\
& \;\;\;\; + (c1_r- \Delta) \otimes (c1_r- \Delta).
\end{align}
\vspace{-.5 cm}
\begin{align}
\hspace{-.4 cm} \text{Therefore,} \;\;\;  &||(\hat{y} \otimes \hat{y}) (t_i)  - (y_s \otimes y_s)(t_i) || \leq 2\norm{y_s(t_i)}\norm{c1_r- \Delta} + \norm{c1_r- \Delta}\norm{c1_r- \Delta},  \nonumber \\
&||(\hat{y} \otimes \hat{y}) (t_i)  - (y_s \otimes y_s)(t_i) || \leq \underbrace{2 ||y_s (t_i)||(c\sqrt{r}+||\Delta||) + (c\sqrt{r}+||\Delta||)^2}_{:= k_1(b,\epsilon)}. 
\end{align}
The upper bounds are written as a function of $b$ and $\epsilon$ to show the dependency. Similarly, it can be shown that, $||(y_s \otimes y_s)(t_j)  - (\hat{y} \otimes \hat{y}) (t_j) || \leq k_1(b,\epsilon)$.
This gives,
$
 ||((\hat{y} \otimes \hat{y}) (t_i) - (\hat{y} \otimes \hat{y}) (t_j) ) - ((y_s \otimes y_s)(t_i) - (y_s \otimes y_s)(t_j)) || \nonumber 
\leq 2k_1(b,\epsilon).
$
Therefore we would have,
\begin{align}
||\delta_{\hat{y}\hat{y}} - \bar{\delta}_{y_s y_s}|| \leq \sqrt{4k_1^2 + ..+ 4k_1^2} := k_2(b,\epsilon).
\end{align}
Assuming $||y_s(t)||$ is bounded for finite time, then for the integral terms we have,

\begin{align}
& \hspace{-.3 cm} \int_{t_1}^{t_1+T}(\hat{y} \otimes \hat{y})d\tau = \int_{t_1}^{t_1+T}(y_s \otimes y_s)d\tau \; + \nonumber \\ & \int_{t_1}^{t_1+T}(y_s(\tau) \otimes
  (c1_r - \Delta)  + (c1_r - \Delta)\otimes y_s(\tau) \; +  (c1_r - \Delta)\otimes (c1_r - \Delta))d\tau .\nonumber
\end{align}
Thereafter we proceed with the following calculations,
\begin{align}
\nonumber
& \norm{\int_{t_1}^{t_1+T}(\hat{y} \otimes \hat{y})d\tau - \int_{t_1}^{t_1+T}(y_s \otimes y_s)d\tau} \leq  \nonumber \\ 
 & \norm{\int_{t_1}^{t_1+T}(y_s(\tau) \otimes (c1_r - \Delta) + (c1_r - \Delta) \otimes y_s(\tau) + (c1_r - \Delta) \otimes (c1_r - \Delta))d\tau}, \nonumber\\
& \leq \int_{t_1}^{t_1+T} (||y_s(\tau)||(c\sqrt{r}+||\Delta||) +(c\sqrt{r}+||\Delta||)||y_s(\tau)||  +(c\sqrt{r}+||\Delta||)^2)d\tau, \nonumber \\ \nonumber
& \leq \int_{t_1}^{t_1+T} \tilde{k}d\tau = T\tilde{k} := k_3(b,\epsilon).\nonumber \\
& \text{i.e.,\;\;} ||I_{\hat{y}\hat{y}} - \bar{I}_{y_sy_s}|| \leq \sqrt{k_3^2+..+k_3^2} := k_4(b,\epsilon).
\end{align}
Similarly assuming $||u_0||$ is bounded for finite time, it can be shown that,
 \begin{align}
  \nonumber
 &\norm{\int_{t_1}^{t_1+T}(  \hat{y} \otimes u_0)d\tau - \int_{t_1}^{t_1+T}(y_s \otimes u_0)d\tau}  \\ \nonumber 
 &\leq \int_{t_1}^{t_1+T} (c\sqrt{r}+||\Delta||)|| u_0(\tau)|| d\tau = T\tilde{k_1} := k_5(b,\epsilon). \nonumber
 \end{align}
 Thereafter, we have
$
 || I_{\hat{y}u_0} - \bar{I}_{y_su_0} || \leq \sqrt{k_5^2+..+k_5^2} := k_6(b,\epsilon).
$
Next we bound the term $||\hat{\Theta}_k - \bar{\Theta}_k||$ as follows.
\begin{align}
&\hat{\Theta}_k - \bar{\Theta}_k = [\delta_{\hat{y}\hat{y}} - \bar{\delta}_{y_sy_s} , -2\bar{I}_{y_sy_s}(I_r \otimes \bar{K}_k^TR) \nonumber \\ & -2\bar{I}_{y_su_0}(I_r \otimes R) +(2I_{\hat{y}\hat{y}}(I_r \otimes K_k^TR) +2I_{\hat{y}u_0}(I_r \otimes R))].
\end{align}
Considering norm of the (1,2) element of $\hat{\Theta}_k - \bar{\Theta}_k$ we have,
\begin{align}
&||(2I_{\hat{y}\hat{y}}(I_r \otimes K_k^TR) +2I_{\hat{y}u_0}(I_r \otimes R))  - 2\bar{I}_{y_sy_s}(I_r \otimes \bar{K}_k^TR) -2\bar{I}_{y_su_0}(I_r \otimes R)|| \nonumber \\
&= ||2(I_{\hat{y}\hat{y}} - \bar{I}_{y_sy_s})(I_r \otimes \bar{K}_k^TR) + 2I_{\hat{y}\hat{y}}(I_r \otimes \Delta K_k^TR) + 2(I_{\hat{y}u_0} - \bar{I}_{y_su_0} )(I_r \otimes R)||.
\end{align}
Now we proceed with the iteration wise analysis.\\
\textit{Iteration - $0$}: We have for the initial stabilizing $K_0$, $\Delta K_0 = 0$. Therefore we can write,
\begin{align}
& ||2(I_{\hat{y}\hat{y}} - \bar{I}_{y_sy_s})(I_r \otimes \bar{K}_0^TR)  + 2(I_{\hat{y}u_0} - \bar{I}_{y_su_0} )(I_r \otimes R)|| \leq \nonumber \\ & \;\;\;\;  2k_4\sqrt{n}||\bar{K}_0^TR|| + 2k_6\sqrt{n}||R|| := k_7(b,\epsilon).
\end{align}
This will give,
\begin{align}
||\hat{\Theta}_0 - \bar{\Theta}_0|| \leq \sqrt{k_2^2 + k_7^2} := k_8(b,\epsilon).
\end{align}
Similarly it can be shown that,
\begin{align}
||\hat{\Phi}_0 - \bar{\Phi}_0|| \leq k_9(b,\epsilon).
\end{align}
Recalling that we have,
\begin{align}
\bar{\Theta}_0 \begin{bmatrix}
vec(\bar{P}_0) \\ vec(\bar{K}_{1})
\end{bmatrix} = \bar{\Phi}_0, \;\;\; \hat{\Theta}_0 \begin{bmatrix}
vec(P_0) \\ vec(K_{1})
\end{bmatrix} = \hat{\Phi}_0.
\end{align}
We can write,
\begin{align}
& \hat{\Theta}_0 \begin{bmatrix}
vec(\Delta P_0) \\ vec(\Delta K_{1})
\end{bmatrix} + (\hat{\Theta}_0 - \bar{\Theta}_0 ) \begin{bmatrix}
vec(\bar{P}_0) \\ vec(\bar{K}_{1})
\end{bmatrix}  = \hat{\Phi}_0 - \bar{\Phi}_0 , \nonumber \\
& \begin{bmatrix}
vec(\Delta P_0) \\ vec(\Delta K_{1})
\end{bmatrix} = \hat{\Theta}_0^{\dagger}((\hat{\Phi}_0 - \bar{\Phi}_0)- (\hat{\Theta}_0 - \bar{\Theta}_0 ) \begin{bmatrix}
vec(\bar{P}_0) \\ vec(\bar{K}_{1})
\end{bmatrix}), \nonumber \\
&  \norm{\begin{bmatrix}
vec(\Delta P_0) \\ vec(\Delta K_{1})
\end{bmatrix}} = \norm{\hat{\Theta}_0^{\dagger}((\hat{\Phi}_0 - \bar{\Phi}_0)- (\hat{\Theta}_0 - \bar{\Theta}_0 ) \begin{bmatrix}
vec(\bar{P}_0) \\ vec(\bar{K}_{1})
\end{bmatrix})} \nonumber \\
& \leq ||\hat{\Theta}_0^{\dagger}||(k_9 + k_8\norm{\begin{bmatrix}
vec(\bar{P}_0) \\ vec(\bar{k}_{1})
\end{bmatrix}}) := k_{10}(b,\epsilon).
\end{align}
As the vectorized form of $K_1 - \bar{K}_1 = \Delta K_1$ is bounded so we can write using matrix Frobenius norm $||K_1 - \bar{K}_1|| \leq k_{11}(b,\epsilon)$.
Then for the iteration 1 we would have $||\Delta K_1||$ is bounded and therefore the previous computation can be similarly done and we can conclude that there exist positive constants $\rho,\rho_1$, dependent on $b,\epsilon$ such that
\begin{align}
||P_k - \bar{P}_{k} || \leq \rho , \;\;\;  ||K_k - \bar{K}_{k} || \leq \rho_1.
\end{align}
This shows that the matrix perturbation terms $\Delta P, \Delta K$ are bounded, and the bound can be made small by reducing the state estimation error bound $b$ for a fixed $\epsilon$. Moreover if the state estimation error $e(t)$ can be made $O(\epsilon)$ then proceeding with the similar computations we will have $P= \bar{P} + O(\epsilon), K=\bar{K} + O(\epsilon)$.\qed
\normalsize
\vspace{-.3 cm}

\section{Proof of Theorem 4:}
We know that $||\hat{y} - y(t)|| \leq b$, therefore by use of slack variables, we can write $\hat{y} = y(t)+ b1_r - \Delta_b(t)$.
Also, we have $||K_{k+1} - \bar{K}_{k+1} || \leq \rho_1,$ implying $K_{k+1} = \bar{K}_{k+1} + \rho_1I - \Delta_{\rho}$.
Let us denote $b_2 = b1_r - \Delta_b(t), \rho_2(b,\epsilon) = \rho_1I - \Delta_{\rho}$. The feedback control is given by, 
$
u = -K_{k+1}\hat{y} = -K_{k+1}(y+b_2)
$
which will make (3) as 
\begin{align}
\dot{y} = A_{11}y + A_{12}z + B_1(-K_{k+1}(y+b_2)),\\
\epsilon \dot{z} = A_{21}y + A_{22}z + B_2(-K_{k+1}(y+b_2)).
\end{align}
We next re-derive the slow subsystem by substituting $\epsilon = 0$. The slow  manifold is given as 
$
z_s = -A_{22}^{-1}A_{21} - B_2K_{k+1}y_s + A_{22}^{-1}B_2K_{k+1}b_2
$.
Therefore, the slow-subsystem dynamics using $K_{k+1} = \bar{K}_{k+1} + \rho_2$ follows as
\begin{align}
\dot{y}_s = (A_s - B_s\bar{K}_{k+1} + \rho_3)y_s + \rho_4(t).
\end{align}
where $A_s = A_{11} - A_{12}A_{22}^{-1}A_{21}, B_s = B_1 -  A_{12}A_{22}^{-1}B_2$,  $\rho_3 = -B_s\rho_2$ and $\rho_4(t)  = - B_s(\bar{K}_{k+1}b_2(t) + \rho_2(t)b_2(t))$.
Here, $\rho_4(t)$ acts as a disturbance to the dynamics: $\dot{y}_s = (A_s - B_s\bar{K}_{k+1} + \rho_3)y_s$. Therefore, we investigate stability by analysing the disturbance-free dynamics. One can consider the dynamics $\dot{y}_s = (A_s - B_s\bar{K}_{k+1} + \rho_3)y_s$ as a perturbed version of the nominal dynamics,
$
\dot{y}_s = (A_s - B_s\bar{K}_{k+1})y_s.
$
Considering a Lyapunov function $V_k(t) = y_s^T \bar{P}_k y_s$,  and computing its time-derive along $\dot{y}_s = (A_s - B_s\bar{K}_{k+1})y_s$, we get
\begin{align}
\dot{V}_k(t) 
   =y_s^T[\bar{P}_k(A_s-B_s\bar K_{k+1}) 
   + (A_s-B_s\bar K_{k+1})^T\bar{P}_k]y_s, \nonumber 
\end{align}
which, using the proof of Theorem $2$, can be shown to reduce to
\vspace{-.3 cm}
\begin{align}\label{eq:final}
\dot{V}_k(t) = -y_s^T[(\bar{K}_k - \bar{K}_{k+1})^TR(\bar{K}_k - \bar{K}_{k+1})]y_s^T  
 -y_s^T[Q + \bar{K}_{k+1}^TR\bar{K}_{k+1}]y_s .
\end{align}
With $Q \succ 0$, closed-loop will be asymptotically stable. The dynamics $\dot{y}_s = (A_s - B_s\bar{K}_{k+1} + \rho_3)y_s$ is basically $\dot{y}_s = (A_s - B_s\bar{K}_{k+1})y_s$ perturbed by $\rho_3 y_s$ vanishing at $y_s=0$. If the estimation error is small with sufficiently small $\epsilon$ then we will have a sufficiently small upper bound $\norm{\rho_3} \leq \bar{\rho}_3$, and the vanishing perturbation $g(t,y_s) = \rho_3 y_s$ will satisfy $||g(t,y_s)|| \leq \bar{\rho}_3 ||y_s||$. With these considerations, we apply~\citet[Lemma 9.1]{khalil} and conclude that the $y_s=0$ is exponentially stable for a sufficiently small $\epsilon$ and state estimation error. Disturbance $\rho_4 (t)$ depends on the state estimation error bound $b$ and the controller gain $K_{k+1}$. With arbitrarily small estimation error, the norm of the disturbance can be bounded by sufficiently small upper-bound $||\rho_4 (t)|| \leq \bar{\rho}_4$.

\normalsize
\vspace{-.5 cm}
\section{Derivation of SP form for clustered network:}
Applying \eqref{similarity} to \eqref{eq:statecompact} we get,
\begin{subequations}\label{eq:sp_ori}
\begin{align}
& \dot{y} = (T_1 \otimes F)Uy + (T_1 \otimes F)G^\dagger z + \epsilon H_{11}y + \epsilon H_{12}z + \tilde{B}_1u,\\
& \dot{z} = (G_1 \otimes F)Uy + (G_1 \otimes F)G^\dagger z + \epsilon H_{21}y + \nonumber \\ & \;\;\;\;\; (H_2 + \epsilon H_{22})z + \tilde{B}_2u,
\end{align}
\end{subequations}

where, $H_{11} = T(L^E \otimes I_s)U, H_{12} = T(L^E \otimes I_s)G^\dagger, H_{21} = G(L^E \otimes I_s)U, H_{2} = G(L^I \otimes I_s)G^\dagger, H_{22} = G(L^E \otimes I_s)G^\dagger, \tilde{B}_1 = TB$, and $\tilde{B}_2 = GB$. 
We can write $U=U_1 \otimes I_s$, where $U_1 = diag(U_{11},\dots,U_{1r}),$ $ U_{1\alpha} = \bf{1}_{n_\alpha}$ for area $\alpha$. We can write $T_1 = N_a^{-1}U_1^T$ where $N_a = diag(n_1,n_2,\dots, n_r)$; $n_\alpha$ is the number of agents in area $\alpha$. Also we have $N_a^{-1}U_1^TU_1 = I_r$, which gives us
\begin{subequations}\label{eq:sp_ori}
\begin{align}
& \dot{y} = (I_r \otimes F)y + (T_1 \otimes F)G^\dagger z + \epsilon H_{11}y + \epsilon H_{12}z + \tilde{B}_1u,\\
& \dot{z} = (G_1U_1 \otimes F)y + (G_1 \otimes F)G^\dagger z + \epsilon H_{21}y + \nonumber \\ & \;\;\;\;\; (H_2 + \epsilon H_{22})z + \tilde{B}_2u,
\end{align}
\end{subequations}

Using the property $(X \otimes Y)^{-1} = X^{-1} \otimes Y^{-1}$, we can write $G^\dagger = G_1^\dagger \otimes I_s$, resulting in
\begin{align}
    (T_1 \otimes F)(G_1^\dagger \otimes I_s) = T_1G_1^{\dagger} \otimes F = 0
\end{align}
Again $G_1U_1 = 0$ from the structures of $G_1$ and $U_1$. Simplifying the term $(G_1 \otimes F)G^\dagger$ we have,
\begin{align}
    (G_1 \otimes F)(G_1^\dagger \otimes I_s) = (G_1G_1^\dagger \otimes F)= (I_{n-r} \otimes F).
\end{align}
Therefore the resultant dynamics becomes  - 
\begin{align}
    & \dot{y} = (I_r \otimes F)y +  \epsilon H_{11}y + \epsilon H_{12}z + \tilde{B}_1u,\\
& \dot{z} =  (I_{n-r} \otimes F)z + \epsilon H_{21}y + (H_2 + \epsilon H_{22})z + \tilde{B}_2u,
\end{align}
In order to get the standard singular perturbation form, we redefine the time scale as $t_s = \epsilon t$, which will lead to \eqref{eq:sp_stan}. \qed
\section{Proof of Lemma 2:}
We consider the actual and the cluster-wise decoupled dynamics as follows
\begin{align}
  & \dot{x} = (I_r \otimes F + (L^I + \epsilon L^E) \otimes I_s)x_d + Bu,\\
&    \dot{x}_d = (I_r \otimes F + L^I \otimes I_s)x + Bu.
\end{align}
Let us denote $F_1 = (I_n \otimes F)$ then we have
\begin{align}
& x(t) = e^{[F_1 + (L^I + \epsilon L^E)\otimes I_s]t}x_0 + \nonumber \\ & \;\;\;\;\;\;\;\;\; \int_{0}^{t} e^{[F_1 + (L^I + \epsilon L^E)\otimes I_s](t - \tau)}B u(\tau)d \tau,\\ 
& x_d(t) = e^{[F_1 + L^I\otimes I_s ]t}x_0 + \int_{0}^{t} e^{[F_1 + L^I\otimes I_s ](t - \tau)}B u(\tau)d \tau. 
\end{align}
\normalsize
Then we can write
\begin{align}
x(t) - x_d(t) = \underbrace{(e^{[F_1 + (L^I + \epsilon L^E)\otimes I_s]t} - e^{[F_1 + L^I\otimes I_s ]t})x_0}_{q_1} + \\ \nonumber
\underbrace{\int_{0}^{t}e^{[F_1 + (L^I + \epsilon L^E)\otimes I_s](t - \tau)} - e^{[F_1 + L^I\otimes I_s ](t - \tau)}B u(\tau)d \tau}_{q2}.
\end{align} 
The norm of $q_1$ can be computed as,
\begin{align}
||q_1|| & = ||((I + [F_1 + (L^I + \epsilon L^E)\otimes I_s]t + \nonumber \\ &([F_1 + (L^I + \epsilon L^E)\otimes I_s]t)^2/2! + \dots)- \\ \nonumber
& \hspace{-.4 cm} (I + (F_1 + L^I\otimes I_s)t + ((F_1 + L^I\otimes I_s)t)^2/2! + \dots))x_0||,\\
& = ||\epsilon[(L^{E} \otimes I_s)t + \dots]x_0|| \leq \epsilon\tilde{k}_1 \;\; \text{for finite }t.
\end{align}
\par
\normalsize
For $||q_2||$, we have
\begin{align}
\hspace{-.4 cm}||q_2|| = \norm{\int_{0}^{t}e^{[F_1 + (L^I + \epsilon L^E)\otimes I_s](t - \tau)} - e^{[F_1 + L^I\otimes I_s ](t - \tau)}B u(\tau)d \tau}\\
\leq \int_{0}^{t} \norm{e^{[F_1 + (L^I + \epsilon L^E)\otimes I_s](t - \tau)} - e^{[F_1 + L^I\otimes I_s ](t - \tau)}B u(\tau)}d \tau.
\end{align}
\normalsize
Computing matrix exponential similarly, we obtain for finite $t \in [0,t_1]$, $||q_2|| \leq \epsilon \tilde{k}_2$. Thus, we conclude
\begin{align}
& \norm{x(t) - x_d(t)} \leq \epsilon (\tilde{k}_1 + \tilde{k}_2) = \epsilon \tilde{k}_3, 
\end{align} 
which means $x(t) = x_d(t) + O(\epsilon)$ for $t \in [0,t_1]$. As $y^\alpha$ and $y_d^\alpha$ are the cluster-wise average variables we conclude the proof. \qed 

\section{Proof of Theorem 7:}
We first show that the learned decentralized control gain $K^\alpha_{k+1}$ can stabilize the decoupled $y_d$ dynamics when $\epsilon$ is small with a sufficiently large $Q$. Then using vanishing perturbation conditions for the reduced slow sub-system dynamics $y_s$, we show that the learned controller will stabilize the actual reduced subsystem, thereby ensuring the overall stability. Let the area-wise control be $u^\alpha = -M^\alpha K_{k+1}^\alpha y^\alpha$. Therefore, $u=-MK_{k+1}y$, where  $K_{k+1}=diag(K_{k+1}^1,\dots,K_{k+1}^r)$.
From Theorem $6$, we have $K_{k+1}^\alpha  = \bar{K}_{k+1}^\alpha + O(\epsilon)$, implying $K_{k+1}  = \bar{K}_{k+1} + O(\epsilon)$. Using the learned gains $K_{k+1}$ for the decoupled dynamics with $F_1 = I_n \otimes F$
%
we get
\begin{align}
& \dot{y}_d = F_1y_d - \tilde{B}_1(-M\bar{K}_{k+1}y_d) -  O(\epsilon)y_d. \label{eq:slow_f}
\end{align}
Next, consider the Lyapunov function $V_k(t) = y_d^T\bar{P}_ky_d$ with $\bar{P}_k \succ 0$, and its time derivative along~\eqref{eq:slow_f} as,
\begin{align}
 \dot{V}_k(t)  &=y_d^T[\bar{P}_k(F_1 -\tilde{B}_1M\bar K_{k+1}-O(\epsilon)) + (F_1-\tilde{B}_1M\bar K_{k+1}-O(\epsilon))^T\bar{P}_k]y_d.
\end{align}



Using the ARE, $A_k^T\bar{P}_k + \bar{P}_kA_k = -(\bar{K}_k^TR\bar{K}_k + Q)$ with $A_k =F_1 -\tilde{B}_1M\bar{K}_k$, and $\bar{K}_{k+1} = R^{-1}M^T\tilde{B}_1^T\bar{P}_k$, it can be shown that
$\dot V_k(t)$ becomes

\begin{align}
\dot V_k(t) &= -y_d^T\left[(\bar{K}_k - \bar{K}_{k+1})^TR(\bar{K}_k - \bar{K}_{k+1})+Q+\bar{K}_{k+1}^TR\bar{K}_{k+1}-O(\epsilon)\right]y_d .
\end{align}
We conclude that with a sufficiently small $\epsilon$, if $ Q=diag(Q^1,\dots,Q^r)$ has a sufficiently large $\lambda_{min}(Q)>0$ then $\dot V_k(t)$ will be negative definite, stabilizing the decoupled dynamics. Next, consider the reduced slow sub-system dynamics of the actual system. Using the learned feedback in \eqref{eq:sp_stan} we get
\begin{align}
& \frac{dy}{dt_s} = ((I_r \otimes F)/\epsilon) y + H_{11}y + H_{12}z + \tilde{B}_1/\epsilon (-MKy),\\
& \epsilon \frac{dz}{dt_s} = \epsilon H_{21}y + (\tilde{H}_2 + \epsilon H_{22})z + \tilde{B}_2(-MKy),
\end{align}
where, $\tilde{H}_2 = H_2 + (I_{n-r}\otimes F)$.
By substituting $\epsilon =0$ and using the slow manifold variable $z_s = -\tilde{H}_2^{-1}\tilde{B_2}(-MKy_s)$, we obtain the reduced sub-system as
\begin{align}
\hspace{-.3 cm}\frac{dy_s}{dt_s} =\frac{(I_r \otimes F)}{\epsilon} y_s+ H_{11}y_s +  (\frac{\tilde{B}_1}{\epsilon} - H_{12}\tilde{H}_2^{-1}\tilde{B}_2)(-MKy_s).\nonumber
\end{align}
Reverting back to the original time-scale $t_s = \epsilon t$, we get
\begin{align}
\hspace{-.4 cm}\frac{dy_s}{dt}  &= (I_r \otimes F) y_s + \epsilon H_{11}y_s +(\tilde{B}_1 - \epsilon H_{12}\tilde{H}_2^{-1}\tilde{B}_2)(-MKy_s),
\nonumber \\ 
 & = (I_r \otimes F)y_s-\tilde{B}_1MK y_s + \epsilon \underbrace{(H_{11} + H_{12}\tilde{H}_2^{-1}\tilde{B}_2)}_{\tilde{H}}y_s. \label{eq:perturbed}
\end{align}

The dynamics \eqref{eq:perturbed} can be viewed as the decoupled dynamics $\dot{y}_d = F_1 y_d -\tilde{B}_1 MKy_d$ perturbed by an $O(\epsilon)$ term vanishing at $y_s=0$. The vanishing perturbation term given by $g(t,y_s) = \epsilon \tilde{H} y_s$ satisfies $\norm{g(t,y_s)} \leq {\epsilon}||\tilde{H}||||y_s||$. With these considerations, we apply~\citet[Lemma 9.1]{khalil} and conclude that $y_s=0$ is exponentially stable for a sufficiently small $\epsilon$. As the slow reduced sub-system model is the perturbed version of the decoupled model with the above-mentioned bound, the learned decentralized controller will exponentially stabilize the slow sub-system dynamics, which in turn stabilizes the entire system with the assumption that the fast sub-system is stable. 
 \qed \par

\end{document}